\documentclass[unnumsec,webpdf,modern,large]{oup-authoring-template}%

\graphicspath{{Fig/}}

\usepackage{csvsimple}
\usepackage{longtable,booktabs}
\usepackage{etoolbox} 
\usepackage{siunitx}
\usepackage[utf8]{inputenc} 

\theoremstyle{thmstyleone}%
\theoremstyle{thmstyletwo}%
\theoremstyle{thmstylethree}%

\begin{document}

\journaltitle{PNAS Nexus}
\DOI{DOI added during production}
\copyrightyear{2026}
\pubyear{2026}
\vol{XX}
\issue{x}
\access{Published: Date added during production}
\appnotes{Paper}

\firstpage{1}


\title[Hyperscaling of spatial fluctuations in urban populations]{Hyperscaling of spatial fluctuations constrains the development of urban populations}

\author[1,2,3,4,$\ast$]{Wout~Merbis\ORCID{0000-0003-3565-0663}}
\author[1,5]{Fernando~A.~N.~Santos\ORCID{0000-0002-6403-8396}}
\author[1,3,6]{Jay~Armas\ORCID{0000-0002-6351-3212}}
\author[1,2,5]{Frank~Pijpers\ORCID{0000-0001-7572-9435}}
\author[1,7]{Mike~Lees\ORCID{0000-0002-5457-9180}}

\address[1]{\orgdiv{Dutch Institute for Emergent Phenomena}, \orgname{University of Amsterdam}, \orgaddress{\street{Oude Turfmarkt 147}, \postcode{1012 GC}, \state{Amsterdam}, \country{the Netherlands}}}
\address[2]{\orgname{Centraal Bureau voor de Statistiek}, \orgaddress{\street{Henri Faasdreef 312}, \postcode{2492 JJ}, \state{Den Haag}, \country{the Netherlands}}}
\address[3]{\orgdiv{Institute for Theoretical Physics}, \orgname{University of Amsterdam}, \orgaddress{\street{Science Park 904}, \postcode{1098 XH}, \state{Amsterdam}, \country{the Netherlands}}}
\address[4]{\orgdiv{Department of Mathematics}, \orgname{Vrije Universiteit Amsterdam}, \orgaddress{\street{Boelelaan 1111}, \postcode{1081 HV}, \state{Amsterdam}, \country{the Netherlands}}}
\address[5]{\orgdiv{Korteweg-de Vries Institute for Mathematics}, \orgname{University of Amsterdam}, \orgaddress{\street{Science Park 105-107}, \postcode{1098 XG}, \state{Amsterdam}, \country{the Netherlands}}}
\address[6]{\orgdiv{The Niels Bohr Institute}, \orgname{University of Copenhagen}, \orgaddress{\street{Blegdamsvej 17}, \postcode{2100}, \state{Copenhagen}, \country{Denmark}}}
\address[7]{\orgdiv{Computational Science Lab}, \orgname{University of Amsterdam}, \orgaddress{\street{Science Park 900}, \postcode{1098 XH}, \state{Amsterdam}, \country{the Netherlands}}}

\corresp[$\ast$]{To whom correspondence should be addressed: \href{email:w.merbis@uva.nl}{w.merbis@uva.nl}}

\received{Date}{0}{Year}
\revised{Date}{0}{Year}
\accepted{Date}{0}{Year}


\abstract{Urban populations exhibit fractal organization and systematic scaling regularities, yet the scaling exponents reported across cities vary substantially, challenging existing theory. Using 100~m gridded population maps for 109 urban regions in the Netherlands (2000--2023) and 368 major world cities (1975--2020), we recursively coarse-grain each city and quantify how the mean and variance of inhabitants in square grid cells of side length $\ell$ scale with $\ell$. This yields two exponents, $\beta$ from $\langle N_\ell\rangle\sim \ell^{\beta}$ and $\gamma$ from $\mathrm{Var}(N_\ell)\sim \ell^{\gamma}$, where in the small-$\ell$ limit $\beta$ equals the planar fractal dimension of populated space. Across cities within a given year, $\gamma$ depends linearly on $\beta$. Compiling $>$10,000 exponent estimates over five decades shows that this hyperscaling relation is robust yet non-universal: its slope and intercept vary across continents and drift systematically in time, trending toward the limiting form $\gamma\simeq 2+\beta$. A mean-field (independent-cell) argument predicts a quadratic mean--variance mapping and cannot reproduce the observed $\beta$--$\gamma$ dependence, implying strong spatial correlations. We derive a correlation-aware variance decomposition in which $\gamma$ is controlled by a correlation dimension $D_c$; in the correlation-dominated regime $\gamma=2+D_c$. If large maturing cities, as are the ones selected in our dataset, evolve to effective monofractal ($D_c\simeq \beta$) cities, the asymptotic prediction becomes $\gamma\simeq 2+\beta$, consistent with the observed temporal drift. This interdependence links urban geometry and fluctuations, provides an empirical constraint for mechanistic models of urban growth, and implies scaling predictions for spatial indicators built from local means and variances.} 

\keywords{Urban scaling laws, Urban growth, Hyperscaling relations}


\otherabstract[Significance statement]{Scaling laws are fundamental to urban science, yet reported exponents vary widely across cities. 
Analyzing high-resolution population grids for 477 urban areas over five decades, we show that two distinct features of urban population organization are systematically related: the fractal geometry of the populated space and the way population fluctuations change with spatial scale. The exponents describing these properties co-vary linearly, and their relationship shifts across continents and over time. We show that spatial correlations between population densities provide a theoretical explanation for this interdependence and predict a limiting relation approached by many large cities. The measured relationship provides an empirical benchmark for urban growth models and constrains scaling predictions for other spatial indicators that depend on population means and variances.}

\maketitle

\section{Introduction}
The dynamical process of urban growth, whether forged as emergent collective behavior of individual entities or shaped by the coordinated action of government officials and city planners, results in fractal-like structures with intricate pattern formation across several scales \cite{mandelbrot1983fractal, batty1994fractal, bettencourt2021introduction}. The geometry of cities has long been linked to the scaling of urbanized areas with the population contained in the urban region \cite{shen2002fractal}. In recent decades, a science of urban scaling laws has emerged, aiming to characterize the efficiencies, benefits, and drawbacks that cities face as they grow \cite{batty2013new}. While household resource consumption and basic stock variables (e.g., housing, waste production) typically scale linearly with city population, physical infrastructure frequently exhibits sublinear scaling. Conversely, socioeconomic outputs—including GDP, wages, innovation, and crime rates—consistently scale superlinearly \cite{bettencourt2007growth,bettencourt2010urban,bettencourt2013origins}. This divergence highlights the increasing ``returns to scale'' inherent to socioeconomic interactions in larger urban agglomerations.

At the same time, a growing body of work has emphasized that the reported scaling exponents are sensitive to how cities are defined and delineated. Estimates of the fractal dimension are known to vary with the spatial scale of analysis and with the distinction between urban cores and peri-urban regions \cite{benguigui2000and,de2003using,lagarias2020comparing,lagarias2021fractal}, and comparative studies across Europe highlight systematic differences linked to regional morphology and data conventions \cite{bettencourt2016urban,van2016urban}. More broadly, several reviews and large-scale empirical assessments have questioned the universality of urban scaling laws, showing that exponents can shift under alternative boundary definitions or across different continents and geographical regions \cite{arcaute2015constructing,cottineau2017diverse,lobo2020urban,ribeiro2024urban,ribeiro2023mathematical,burger2022global,hendrick2025stochastic, depersin2018global,prieto2025violence}. Recent work on urban regions using tools from surface growth physics report non-universal values for exponents controlling growth and correlation length \cite{marquis2025universal} (whereas the local roughness exponent characterizing urban surface irregularities does appear to be universal). These findings do not invalidate scaling theory, but instead suggest that scaling relations may encode underlying spatial organization in ways that are not yet fully understood.

Although many urban indicators scale with the total population, much less is known about how \emph{spatial fluctuations} in population depend on the underlying multiscale organization of cities. In this paper we analyze gridded population data using a coarse-graining procedure that aggregates inhabitants within square regions of linear size $\ell$, and we quantify how both the mean and the variance of these coarse-grained counts scale with $\ell$. This yields two exponents, $\langle N_\ell \rangle \sim \ell^{\beta}$ and ${\rm Var}(N_\ell) \sim \ell^{\gamma}$. In the small-$\ell$ limit, $\beta$ coincides with the planar fractal dimension $d_f$ of urban occupancy. 
Our central empirical observation is that the fluctuation exponent $\gamma$ co-varies systematically with the fractal dimension $\beta$: across cities within a given year, the exponent pair $(\beta,\gamma)$ lies close to a line. 
We refer to this interdependence as a hyperscaling relation. The terminology is borrowed from statistical physics, where scaling exponents are often linked by nontrivial relations rather than varying independently \cite{widom1965surface,stanley1971phase}. We use the term here in a broader sense to emphasize that the two urban scaling exponents are empirically constrained relative to one another. As shown later, the theoretical interpretation of this relation involves the spatial dimensionality of the population field, providing a closer connection to the conventional statistical-physics notion of hyperscaling.\footnote{The term “hyperscaling” is also used in organization and strategy research to describe the exceptional scalability of digital firms and platforms \cite{giustiziero2023hyperspecialization}. In that literature, hyperscaling refers to an organizational growth logic arising from highly scalable digital resources and is not an exponent identity. Our use instead follows statistical physics, where hyperscaling denotes relations among scaling exponents that explicitly involve the spatial dimension.}
Similar empirical relations have recently been reported in other complex systems such as the brain \cite{castro2025interdependent}.

\begin{figure*}[!t]
 \centering
 \includegraphics[width=\textwidth]{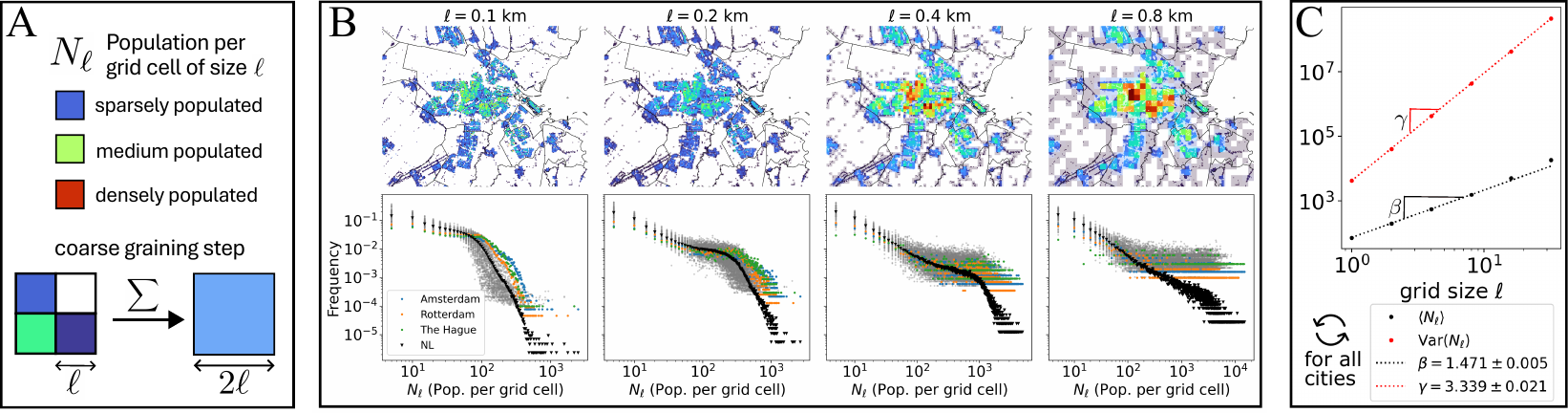}
 \caption{\textbf{Scaling analysis pipeline for grid-level population data.}
 (A) Coarse-graining the population data by doubling the box size $\ell$ and aggregating inhabitants in each box. Colors indicate the number of residents within boxes of area $\ell^2$.
 (B) Top: example coarse-graining steps for Amsterdam, with administrative boundaries shown in black. Bottom: empirical distributions of $N_\ell$ for four values of $\ell$; grey points show city-level estimates, and black markers summarize the Netherlands-wide aggregation.
 (C) Mean and variance of $N_\ell$ for the city of Amsterdam as a function of $\ell$ in log--log coordinates; slopes define $\beta$ and $\gamma$ (Eq.~\ref{eq:beta_gamma_def}). Data shown here are for the year 2023 \cite{cbs2024dataset}. \textcopyright~Centraal Bureau voor de Statistiek, \textcopyright~ESRI Nederland.}
 \label{fig:renormalizeCBS}
\end{figure*}

At first sight, the coexistence of power-law scaling in the mean and in the variance may evoke Taylor’s law \cite{taylor1961aggregation}, which describes empirical power-law relations between variance and mean in ecological populations. However, our result is not a trivial example of Taylor’s law. Rather than relating variance to mean across an ensemble of independent systems, we analyze how both quantities scale with spatial coarse-graining within the same urban system. The observed interdependence between $\beta$ and $\gamma$ therefore constitutes a multi-scale generalization of variance–mean scaling, in which the exponents governing spatial aggregation are themselves constrained by the underlying fractal geometry and spatial correlations of the urban population.

Our approach also connects to statistical-physics-inspired models of urban form. Correlated percolation models of urban growth \cite{makse1998modeling} demonstrate how spatial correlations and density gradients can generate fractal urban patterns, while algorithmic clustering procedures such as the City Clustering Algorithm \cite{rozenfeld2008laws} show that scaling relations are sensitive to how spatial aggregation is defined. Similarly, spatial network models emphasize how transport constraints and spatial embedding shape urban morphology and scaling behavior \cite{louf2014scaling,louf2014congestion, prietocuriel2023scaling}. In contrast to these generative or delineation-based frameworks, our analysis does not assume a specific growth mechanism or clustering rule; instead, it derives a geometric constraint linking fluctuation scaling to fractal structure. In this sense, the hyperscaling relation we observe can be viewed as a coarse-grained structural signature that any spatially correlated urban growth model must reproduce.

We establish the hyperscaling relation using two complementary datasets. For the Netherlands, we analyze $100\,{\rm m}\times100\,{\rm m}$ population grids provided by Statistics Netherlands (CBS) across 109 urban regions and 24 years (2000--2023) \cite{cbs2024dataset}. Globally, we apply the same pipeline to major cities on multiple continents using the GHS-POP data set (1975--2020) \cite{ghspop2023dataset, Pesaresi2024advances}. Together, these sources yield more than $10^4$ estimated exponent pairs across five decades. This broad sampling reveals that the hyperscaling relation is robust yet non-universal: its slope and intercept vary across world regions and drift over time, indicating that scaling variability is structured rather than arbitrary.

To interpret these patterns, we analyze how spatial coarse-graining acts on the moments of the population field. When a city is aggregated into blocks of size $\ell$, the coarse-grained variance $\mathrm{Var}(N_\ell)$ decomposes into the sum of local cell variances plus a sum of all pairwise spatial covariances $\mathrm{Cov}(X_i,X_j)$ between subcells. A standard ``mean-field'' null baseline assumes fine-grid cells fluctuate independently. This assumption leads to a quadratic mean--variance relationship that bounds the variance exponent $(\gamma \le 2\beta)$ and fails to capture the empirical co-variation of $\beta$ and $\gamma$. By contrast, explicitly accounting for spatial dependence between population values in different locations shows that the variance scaling can become dominated by the covariance contribution. This term is governed by a spatial correlation dimension $D_c$, yielding $\gamma = 2 + D_c$ in the correlation-dominated regime and approaching $\gamma \simeq 2+\beta$ when the urban population approaches a monofractal spatial distribution ($D_c \simeq \beta$).

Across our global sample of major metropolitan areas, the empirical hyperscaling relation drifts systematically toward this theoretically predicted limiting form over the 1975--2020 observation period. This agreement is consistent with an effective transition toward correlation-dominated, monofractal spatial organization in large, rapidly urbanizing systems. This convergence strongly aligns with independent longitudinal observations of London's street network, which similarly documented a transition from a multifractal to a monofractal topology over time \cite{murcio2015multifractal}.

Beyond explaining exponent interdependence, this framework has practical implications: once the joint scaling of mean and variance is constrained, any spatial indicator that depends on these moments (or functions of them) inherits predictable scaling behavior. We therefore view the measured hyperscaling relations as empirical constraints on urban growth processes and as a starting point for relating multiscale population organization to downstream socio-economic and planning-relevant metrics.

\section{Results}\label{sec:results}

\subsection{Mean and Variance scaling exponents in Dutch cities}
We first focus on population scaling within and across cities in the Netherlands, using grid-level population data from Statistics Netherlands (CBS). The dataset reports the number of inhabitants per $100\times100$~m cell for each year from 2000--2023 \cite{cbs2024dataset}. We select 109 urban regions spanning the full range of official urbanization levels (see Appendix \ref{sec:mat&met} for region selection, offsets, and robustness checks). For each region and year, we recursively coarse-grain the population field by aggregating the base grid into square boxes of side
$
\ell \in \{100,200,400,800,1600,3200\}\ \,
$
(in meters), and we compute the distribution of coarse-grained counts $N_\ell$ (Fig.~\ref{fig:renormalizeCBS}A,B). 

Throughout, we focus on \emph{non-empty} boxes to characterize populated urban form and to reduce sensitivity to uninhabitable areas (water bodies, parks, industrial land) and to bounding-box geometry. This restriction also removes a trivial geometric contribution to the mean scaling: if all boxes in a fixed bounding region were included, the mean population per box would scale as the total population divided by the total number of boxes, giving the Euclidean behavior $\langle N_\ell\rangle \sim \ell^2$ by construction.

For each region, we quantify how the first two moments scale with $\ell$,
\begin{equation}
\langle N_\ell\rangle \sim \ell^{\beta},
\qquad
{\rm Var}(N_\ell) \sim \ell^{\gamma},
\label{eq:beta_gamma_def}
\end{equation}
where the exponents $\beta$ and $\gamma$ are estimated from log--log regressions over the small-$\ell$ range (Fig.~\ref{fig:renormalizeCBS}C; see Appendix \ref{sec:mat&met} for fit windows, sensitivity analysis and uncertainty estimates). For the non-empty boxes, the scaling of the mean has a direct box-counting interpretation. Since the total population $N$ of a region is conserved under coarse-graining,
\begin{equation}
\langle N_\ell\rangle
=
\frac{N}{n_{\Box}(\ell)},
\end{equation}
where $n_{\Box}(\ell)$ is the number of non-empty boxes at scale $\ell$. If the populated support is fractal at small scales, $n_{\Box}(\ell)\sim \ell^{-d_f}$, and therefore
\begin{equation}
\langle N_\ell\rangle \sim \ell^{d_f},
\end{equation}
so that $\beta \to d_f$ when the scaling is well developed. Thus, restricting to occupied boxes removes the trivial Euclidean scaling associated with averaging over all boxes and allows $\beta$ to probe the geometry of the populated urban support.

\subsection{An interdependent (hyperscaling) relation between $\beta$ and $\gamma$}
\begin{figure}
    \centering
    \includegraphics[width=\linewidth]{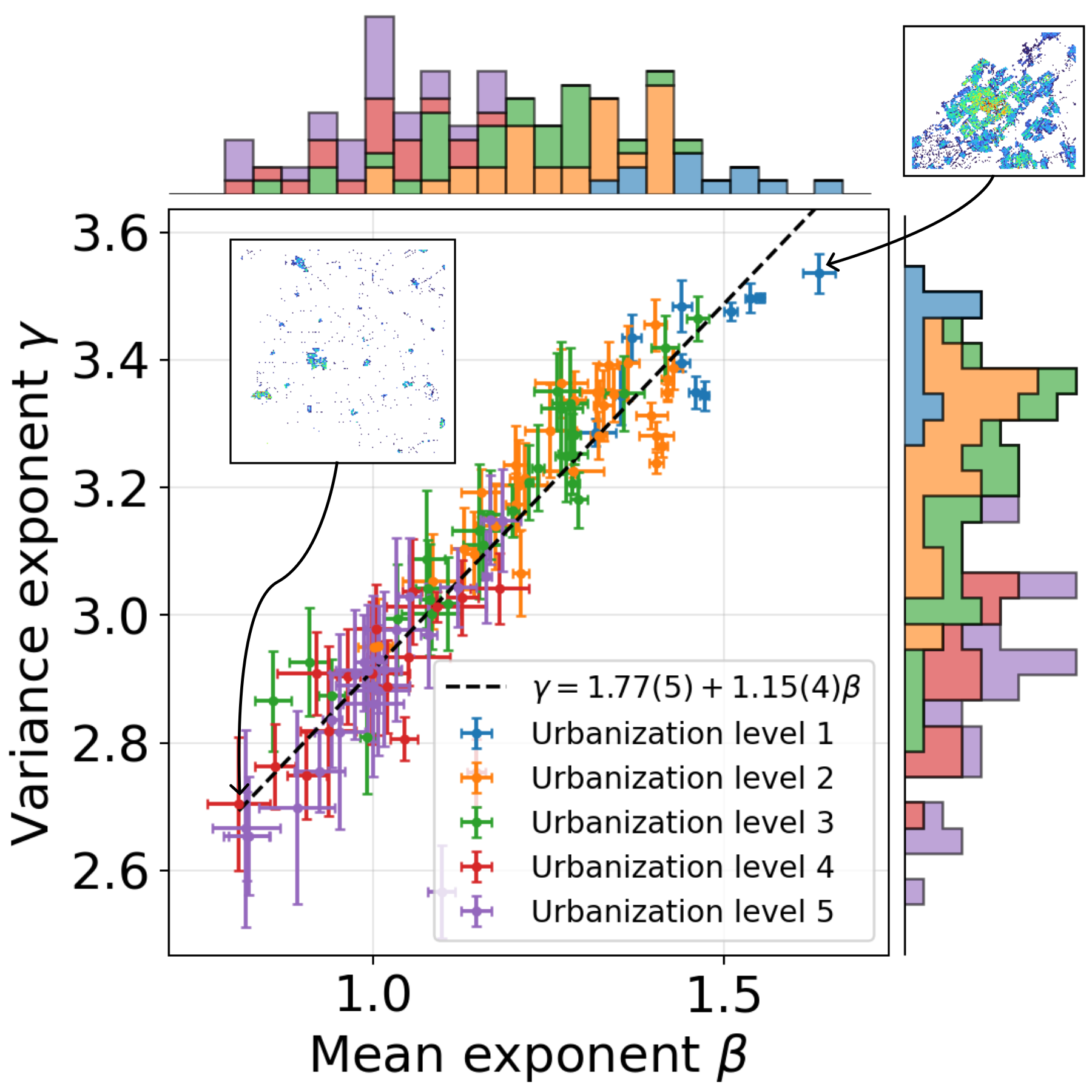}
    \caption{\textbf{Scaling exponents for regions in the Netherlands.}
    Exponents for the mean ($\beta$) and variance ($\gamma$) of inhabitants per box across Dutch urban regions (2023) show a strong linear relationship (Eq.~\ref{eq:hyperscaling_line}). Points are colored by CBS urbanization level. Insets show two representative regions spanning the spectrum (low-density/low-$\beta$ vs.\ high-density/high-$\beta$). Error bars indicate uncertainties from the log--log fits used to estimate $\beta$ and $\gamma$.}
    \label{fig:scaling_analysis}
\end{figure}

Across urban regions in the Netherlands in a fixed year, the exponents $\beta$ and $\gamma$ are not independent. Figure~\ref{fig:scaling_analysis} shows the 2023 exponent pairs: cities align closely along a line,
\begin{equation}
\gamma = c_1 + c_2\,\beta,
\label{eq:hyperscaling_line}
\end{equation}
with a fit that is substantially tighter than would be expected if $\beta$ and $\gamma$ varied independently. Importantly, the $\beta$--$\gamma$ plane also reveals an emergent ordering by development level. When points are colored by CBS urbanization class (see Table \ref{tab:urbanization_levels} in Appendix \ref{app:datasources} for definitions), the classes separate systematically along the hyperscaling line: highly urbanized regions tend to occupy the upper-right part of the plot (larger $\beta$ and larger $\gamma$), whereas weakly urbanized regions lie toward the lower-left. As the urbanization level is not used to estimate $\beta$ or $\gamma$, this separation is not imposed by construction. Rather, it indicates that the multiscale population organization encodes interpretable information about urban development and density. The insets in Fig.~\ref{fig:scaling_analysis} illustrate two examples that span the range, from a sparse and weakly urbanized municipality (Noordoostpolder, low $\beta$) to a dense and highly urbanized municipality (the Hague, high $\beta$).

To test the robustness of the interdependence relations and quantify temporal variability, we repeat the full analysis for each year from 2000--2023. For each year, we find that Eq.~\ref{eq:hyperscaling_line} provides a good description of the cross-sectional relationship between $\beta$ and $\gamma$, but the fitted coefficients $(c_1(t),c_2(t))$ vary over time beyond regression uncertainties. This supports the idea that scaling variability is structured: the relationship between urban form ($\beta$) and fluctuations ($\gamma$) is persistent, while its parameters change over time. The time series of $(c_1,c_2)$, along with the sensitivity analysis (fit windows, offsets, and region selection), are reported in Appendix \ref{sec:mat&met}.

\begin{figure*}
    \centering
    \includegraphics[width=.9\linewidth]{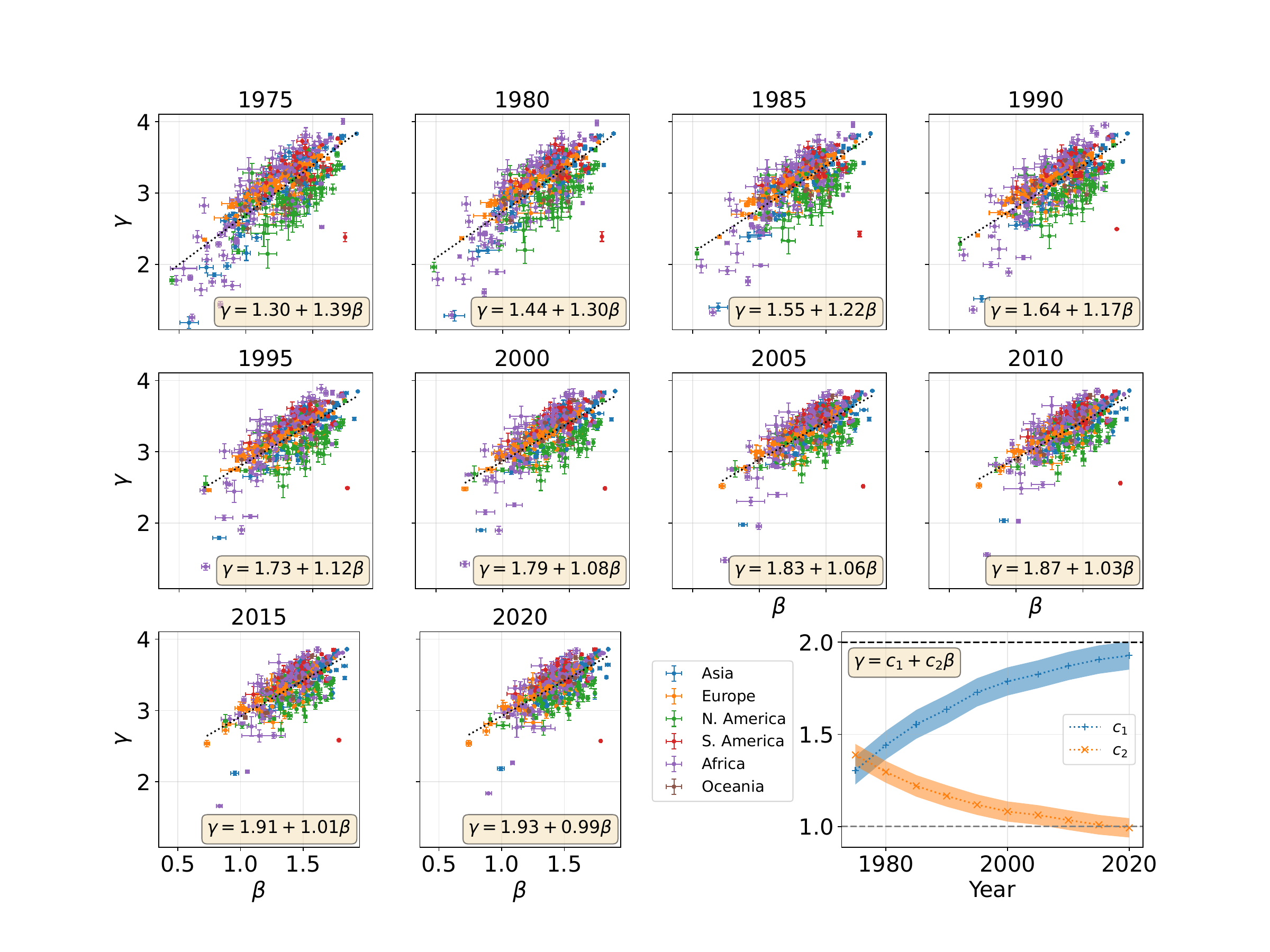}
    \caption{ \textbf{Global hyperscaling across five decades (GHS-POP).}
    Exponent pairs $(\beta,\gamma)$ for major world cities computed from GHS-POP at 5-year intervals from 1975--2020, pooled across cities (see Supplementary Information for continent-level stratification). The selection of cities is kept constant for each year. As cities progress in their evolution, the scaling relationship between $\gamma$ and $\beta$ shows a drift toward $c_1\to2$ and $c_2\to1$, i.e.\ $\gamma\simeq2+\beta$.}
    \label{fig:world_hyperscaling}
\end{figure*}

\subsection{Global hyperscaling and drift toward an asymptotic form}
We next ask whether the hyperscaling relation extends beyond the Netherlands and how it evolves over longer time horizons. We apply the same coarse-graining pipeline to gridded population estimates from the Global Human Settlement Population Data Set (GHS-POP) \cite{ghspop2023dataset, Pesaresi2024advances}. We analyze a curated set of 368 major cities across continents and compute exponents $\beta$ and $\gamma$ at 5-year intervals (dataset time resolution) from 1975--2020. 
The cities were selected to provide broad country-level geographic coverage while retaining sufficiently large, non-overlapping urban regions for the multiscale analysis; full selection criteria and the complete city list are given in Appendix \ref{app:regionselection} and \ref{app:tables}.
Across these decades, we obtain on the order of $10^4$ individual exponent estimates (counting both $\beta$ and $\gamma$ across city-years). As in the Netherlands, each time slice exhibits an approximately linear relationship between $\beta$ and $\gamma$ when cities are pooled (Fig.~\ref{fig:world_hyperscaling}). However, the fitted coefficients depend on time and (more weakly) on region; continent-specific fits are reported in Appendix \ref{app:temp_hyperscaling}.

A key temporal pattern emerges in the global record: the regression coefficients in Eq.~\ref{eq:hyperscaling_line} drift toward
\begin{equation}
c_1 \rightarrow 2,
\qquad
c_2 \rightarrow 1,
\end{equation}
so that the hyperscaling relation approaches the asymptotic form $\gamma \simeq 2+\beta$ (see the bottom right panel of Fig.~\ref{fig:world_hyperscaling}). This trend is consistent with the interpretation that, as cities expand and mature, their multiscale fluctuation structure becomes increasingly constrained by the geometry of their populated support, as we will now demonstrate.

\subsection{Mean-field baseline and a correlation-aware explanation}
To interpret the observed exponent interdependence, we compare against a mean-field baseline in which fine-grid cells contribute independently with finite variance. In that case, coarse-graining implies a \emph{quadratic} mean--variance relationship of Taylor type,
${\rm Var}(N_\ell)\;=\;a\,\langle N_\ell\rangle \;+\; b\,\langle N_\ell\rangle^2$, (see Appendix \ref{app:meanfield} for details),
which maps to effective exponents that do not reproduce the observed linear $\beta$--$\gamma$ relation. Thus, a description that neglects statistical dependence between population values in different cells within the same urban region at the same time cannot account for the observed fluctuation scaling.

Motivated by this, we derive a variance decomposition that separates an ``independent'' term from a covariance-driven term. Writing $N_\ell=\sum_{i\in A_\ell}X_i$ as the sum of fine-grid populations in a box $A_\ell$,
\begin{equation}
{\rm Var}(N_\ell)=\underbrace{\sum_{i\in A_\ell}{\rm Var}(X_i)}_{V_1(\ell)}
\;+\;
\underbrace{\sum_{i\neq j\in A_\ell}{\rm Cov}(X_i,X_j)}_{V_2(\ell)}.
\label{eq:var_decomp}
\end{equation}
Under mild conditions, $V_1(\ell)\propto \ell^2$, while $V_2(\ell)$ is controlled by the spatial covariance function. If covariances decay as a power law with exponent $\eta$, then $V_2(\ell)\sim \ell^{2+D_c}$, where $D_c=2-\eta$ acts as a \emph{correlation dimension}. This yields the scaling form
\begin{equation}
{\rm Var}(N_\ell)=A\,\ell^2 + B\,\ell^{2+D_c},
\label{eq:two_term_var}
\end{equation}
implying an effective variance exponent $\gamma$ that interpolates between $2$ and $2+D_c$ depending on the relative weight of the correlation term (see SI for a detailed derivation). In the correlation-dominated regime, one expects $\gamma=2+D_c$. If maturing urban systems become increasingly monofractal so that $D_c\simeq \beta$, then Eq.~\ref{eq:two_term_var} predicts the asymptotic hyperscaling form $\gamma\simeq 2+\beta$, providing a mechanistic explanation for the temporal drift observed in the global record. Full derivations, assumptions, and regime analyses are provided in the Appendix \ref{app:correlations}.

To further probe the origin of the hyperscaling relation, we analyzed two stylized urban growth models: a correlated percolation model and a dynamical birth–death model with nearest-neighbor interactions (see Appendix \ref{app:models}). In both cases, artificial cities generated under a tunable correlation strength exhibit a clear interdependence between the fractal dimension $\beta$ and the variance exponent $\gamma$. Although these models do not quantitatively reproduce the empirical slopes and intercepts observed in real cities, they qualitatively match the central feature of the data: stronger spatial correlations systematically steepen the $\beta-\gamma$ relation. In contrast, in the weak-correlation limit the relation flattens. These results demonstrate that in addition to density and fractal geometry, the spatial correlation structure is essential to obtain the observed hyperscaling behavior, and they support our interpretation of the empirical drift as a transition toward increasingly correlation-dominated urban organization.

\section{Discussion}\label{sec:discussion}

Our analysis reveals a robust but non-universal interdependence between urban geometry and multiscale fluctuations in population. Across Dutch municipalities (2000--2023) and a global sample of major cities (1975--2020), the mean coarse-grained population obeys $\langle N_\ell\rangle \sim \ell^{\beta}$ and the variance obeys $\mathrm{Var}(N_\ell)\sim \ell^{\gamma}$, with the exponent pair $(\beta,\gamma)$ aligning closely along an approximately linear ``hyperscaling'' relation $\gamma \approx c_1 + c_2 \beta$ at each time slice. Over five decades, the fitted coefficients drift systematically towards $(c_1,c_2)\to (2,1)$, consistent with an asymptotic approach to $\gamma \simeq 2+\beta$. A mean-field (independent-cell) baseline predicts a Taylor-type quadratic mean--variance mapping and does not reproduce the observed $\beta$--$\gamma$ co-variation, whereas a correlation-aware variance decomposition shows that $\gamma$ is controlled by a correlation dimension $D_c$ and, in the correlation-dominated regime, approaches $\gamma = 2 + D_c$; under monofractality hypothesis ($D_c \simeq \beta$), this yields the limiting form $\gamma \simeq 2+\beta$.

It is important to distinguish the different levels of analysis involved in these results. The exponents $\beta$ and $\gamma$ are estimated from spatial coarse-graining \emph{within} each urban region at a given observation time. The hyperscaling relation is then identified primarily through \emph{cross-sectional} comparison of these exponent pairs across different cities within the same year. Finally, the repeated observations allow us to examine how individual cities move through the $(\beta,\gamma)$ plane over time. These intra-urban, inter-urban, and longitudinal comparisons provide complementary information, but they should not be interpreted as interchangeable scaling relations.

Placing these findings in context, the observed non-universality of $(c_1,c_2)$ across continents and years points to the central role of spatial correlations and their origins. In our framework, the geometry of the occupied space controls $\beta$, but the strength and range of spatial covariance determine how rapidly the covariance contribution dominates the variance and thus how large $\gamma$ becomes at fixed $\beta$. This suggests several possible sources of regional variability, including geography and topography (e.g. coastlines, waterways, and terrain constraints), historically contingent growth trajectories (e.g. zoning, transport investments, and industrial development), and institutional differences in planning regimes that shape spatial segregation, polycentricity, and land-use mixing. These factors may influence the population geometry and spatial covariance structure measured by $\beta$ and $\gamma$, but our analysis does not identify their individual causal contributions. The hyperscaling relation should therefore be regarded primarily as a descriptive diagnostic of multiscale population organization. Establishing whether particular planning interventions or infrastructural changes systematically modify this relation requires spatially resolved longitudinal analyses that connect such changes directly to the evolving population field.

Transportation infrastructure provides one concrete example of a process that could alter urban population organization over time. New transport links may reshape the occupied population support measured by $\beta$, while changes in accessibility and connectivity may also affect the spatial covariance structure underlying $\gamma$. To examine whether transport expansion could account for the observed temporal evolution, we performed an exploratory comparison between the longitudinal exponent trajectories of Shanghai and São Paulo with the historical growth of their metro systems (see Appendix \ref{app:metro} and Fig.~\ref{fig:transport_two_city}). The comparison does not support a simple relation in which increasing metro size drives convergence toward the limiting form ($\gamma=2+\beta$). In Shanghai, substantial movement toward this line already occurred before the metro opened, while the rapid network expansion after the mid-1990s was accompanied by little further change in the residual $m=\gamma-(2+\beta)$. A similar pattern is visible in São Paulo: despite continued network growth, $m$ remains nearly constant after approximately 1995.

More generally, these cases illustrate that the two exponents can evolve substantially while remaining tightly coupled within the same city. From the mid-1990s onward, Shanghai and São Paulo follow approximately parallel trajectories in the $(\beta,\gamma)$ plane, with $\gamma-\beta$ remaining nearly constant. Thus large changes in urban form and infrastructure need not erase a city's relative position with respect to the hyperscaling relation. Aggregate transport-network expansion is therefore not sufficient to explain either convergence toward $\gamma=2+\beta$ or the persistent heterogeneity between urban systems. Transportation may nevertheless affect both exponents jointly through spatially heterogeneous changes in accessibility, land use, and residential location. Testing such mechanisms would require spatially resolved measures of transport exposure together with direct measurements of the population covariance structure, rather than aggregate route length alone.

The observed non-universality of $(c_1,c_2)$ across continents and decades emphasizes that the temporal patterns reported here should not be interpreted as a universal trajectory of urban evolution. Our observations cover a specific historical window (2000--2023 for the Dutch data and 1975--2020 for the global sample) and different regions occupy different parts of the ($\beta,\gamma)$ plane and follow different temporal trajectories within that window. In particular, European cities remain comparatively close to the limiting relation $\gamma \simeq 2+\beta$, while cities in Asia and Africa show a more pronounced drift toward it, and cities in North and South America display systematically different fitted slope--intercept combinations (see Fig.~\ref{fig:hyperscale_parameters}). These regional differences may reflect variation in urban morphology, planning institutions, land-use regulation, historical development pathways, and geographical constraints, but the present analysis does not identify the contribution of these factors. Explaining the observed heterogeneity will require combining the scaling analysis with independent measures of urban form, planning context, infrastructure, and spatial population structure.

Several methodological constraints of the current approach warrant consideration. 
First, exponent estimation can in principle be sensitive to the finite range of scales considered and to the delineation of urban regions. We explicitly quantify the former by repeating the analysis over a dense set of linearly spaced coarse-graining scales and multiple nested fitting windows (Appendix~\ref{sec:exponent_robustness}), finding that the estimated exponents and their cross-city relation are generally robust, while also revealing a systematic crossover when larger spatial scales are included. Sensitivity to alternative definitions and delineations of urban regions remains an important direction for future work. Second, data-specific preprocessing can obscure small-$\ell$ behavior: CBS disclosure protocols (such as rounding and low-count suppression) and GHS-POP estimation artifacts may introduce variance biases, thereby affecting $\beta$ and $\gamma$ in sparsely populated zones. Third, our correlation-aware theory assumes approximate stationarity and power-law covariance scaling; directly measuring empirical covariance decay and correlation dimensions would yield more stringent tests of these assumptions. Finally, the stylized growth models detailed in Appendix \ref{app:models} exhibit qualitative alignment but utilize simplifications (e.g., restricted cell capacities) that can distort asymptotic behavior as $\beta \to 2$. Future quantitative comparisons will necessitate richer models equipped with variable capacities and calibrated interaction kernels \cite{marquis2025modeling}.

A key implication of our findings is that any spatial indicator constructed from coarse-grained means and variances is not free to scale arbitrarily: once $(\beta,\gamma)$ are constrained by hyperscaling, downstream quantities depending on these moments inherit restricted scaling behavior. These predictions concern indicators constructed from spatial variation within an urban region and should be distinguished from conventional urban scaling relations obtained by comparing aggregate quantities across cities of different population sizes. This opens the possibility of exploring these constraints for other spatial datasets at pixel level beyond population. A simple follow-up application that would require spatial income distribution is spatial inequality measured through a scale-dependent Gini coefficient, for which broad classes of distributions yield $G_\ell \propto \sigma_\ell/\mu_\ell$ and therefore a Gini exponent $g = \gamma/2 - \beta$, implying feasible bands in the $(\beta,g)$ plane and a direct link between urban geometry and the scaling of inequality. More generally, metrics of specialization, diversity, and functional mixing that depend on heterogeneous spatial concentrations should be jointly constrained by the same correlation structure that governs $\gamma$. This perspective aligns naturally with recent scaling theory emphasizing how diversity and specialization co-vary across complex systems \cite{yang2026scaling}: in cities, multiscale covariance provides a concrete geometric--statistical channel through which the organization of activities and the variability of local intensities become coupled. If validated across additional indicators (e.g., land-use mixtures, accessibility, emissions, or service provision), these constraints could inform comparative urban diagnostics and help distinguish policy-relevant mechanisms (connectivity, zoning, segregation) from purely size-driven effects.

A potentially important extension of this framework concerns conventional socioeconomic urban scaling. Classical urban scaling studies compare aggregate quantities such as economic output, innovation, or crime across cities of different total population, whereas the exponents studied here characterize how population is organized across spatial scales \emph{within} individual cities. These are distinct levels of analysis, and our results do not establish a direct mapping between them. Nevertheless, there is a natural mathematical route by which intra-urban population structure could influence inter-urban scaling. If a socioeconomic output depends nonlinearly and convexly on local population density or interaction intensity, then its city-wide aggregate depends not only on the mean local population but also on its spatial heterogeneity. In particular, an expansion around the mean introduces the spatial variance as the leading correction. Because the scale dependence of population variance is governed by $\gamma$, the hyperscaling relation between $\beta$ and $\gamma$ could therefore place constraints on how such locally generated quantities aggregate at the city level. This suggests a possible connection between intra-urban fluctuation scaling and the variability or deviations observed in conventional cross-city socioeconomic scaling relations. We do not test this connection here, but it provides a concrete direction for future work using spatially resolved socioeconomic data to determine whether differences in internal population organization help explain differences in aggregate urban outputs.

The role of the hyperscaling relation as a constraint should be understood in an inferential rather than a causal sense. In our framework, urban growth processes generate spatial organization and correlations, which in turn determine the observed exponents $\beta$ and $\gamma$. The hyperscaling relation is therefore an emergent consequence of those processes, not a rule that causally directs urban development. At the same time, different mechanisms need not produce the same exponent relation. This is already apparent in our analysis: the independent-cell approximation fails to reproduce the empirical $(\beta, \gamma)$ dependence, while the relations generated by spatially correlated growth models vary with their correlation structure. The empirical hyperscaling relation can therefore be used in reverse as a model-selection criterion: mechanistic models of urban growth should reproduce not only the fractal geometry of populated space, but also the associated multiscale fluctuation structure. In this sense, the measured relation reduces the space of growth mechanisms compatible with the observations.

In closing, our results suggest that variability in reported urban scaling exponents is not merely noise, but structured information about spatial organization. The empirically robust yet drifting hyperscaling relation provides a compact summary of how urban geometry and fluctuations are coupled across cities, how this coupling changes over time, and it provides a compact empirical benchmark for comparing mechanistic models of urban growth. By elevating spatial correlations to a first-class ingredient of urban scaling, this framework opens a path toward connecting multiscale population organization to inequity, functionality, and planning-relevant outcomes within a unified, testable scaling theory.

\section{Acknowledgments}
The authors thank the anonymous reviewers for their valuable suggestions. The authors wish to thank Tuan Minh Pham and Clelia de Mulatier for helpful discussions.
The authors acknowledge and thank the participants of the workshop ``Scaling and Criticality in Complex Systems" for their insightful comments on this work, in particular Geoffrey West, Marc Barthelemy, Tiziana Di Matteo and Miguel Angel Mu\~noz.

\section{Competing interest}
The authors declare that they have no competing interests.

\section{Funding}
W.M. is funded by the NWA-ORC 2023 programme \textit{Emergence at all scales} (EAAS).
F.S. is supported by the Dutch Institute for Emergent Phenomena (DIEP) cluster at the University of Amsterdam under the Research Priority Area Emergent Phenomena in Society: Polarization, Segregation and Inequality.
J.A. is partly funded by the DIEP cluster at the University of Amsterdam via the DIEP programme Foundations and Applications of Emergence (FAEME) and the national NWA-ORC 2023 consortium EAAS.

\section{Author contributions}

W.M. and F.S. conceptualized the project, W.M. performed the data analysis and made all figures, W.M. and F.S. performed the mean-field and spatially correlated fluctuations analysis, W.M. simulated and analyzed urban growth models. W.M., F.S., and M.L. wrote the manuscript. All authors participated in discussions on the interpretation and analysis of the results and revised the manuscript accordingly. All authors edited and reviewed the final manuscript. 

\section{Data availability}
The raw data used in this work is publicly available at \cite{cbs2024dataset,ghspop2023dataset}. All code and scripts used to process the data is made available in a \href{https://github.com/wmerbis/renormalize_the_world}{public GitHub repository}.



\begin{appendices}

\section{Materials and Methods}\label{sec:mat&met}

\subsection{Data sources and description}\label{app:datasources}

Statistics Netherlands (CBS) publishes population statistics at multiple geospatial levels derived from aggregated administrative register data. The most fine-grained publicly available dataset reports the number of inhabitants on a $100 \times 100\ m$ population grid for the years 2000–2024 \cite{cbs2024dataset}. For privacy protection, population counts are reported in multiples of five, and grid cells containing fewer than five individuals (or otherwise sensitive information) are suppressed. Additional statistical disclosure procedures may also have been applied to the dataset \cite{cbsdatabeschrijving}. The urbanization levels reported in Fig.~\ref{fig:scaling_analysis} follow the classification thresholds defined by Statistics Netherlands (Table~\ref{tab:urbanization_levels}).

For the global analysis of major cities we use the GHS-POP R2023A gridded population product from the Copernicus/EC-JRC Global Human Settlement Layer (GHSL) \cite{ghspop2023dataset,Pesaresi2024advances}. GHS-POP provides global residential population estimates per $100\times100\ m$ cell for multiple epochs (1975–2020 in five-year intervals), matching the spatial resolution of the CBS dataset. The dataset combines national and subnational population totals from the Gridded Population of the World (GPW v4.11, CIESIN) with Earth-observation–derived settlement layers. Population counts are disaggregated to fine spatial grids using a dasymetric mapping approach that allocates inhabitants preferentially to built-up areas identified from multi-temporal Landsat and Sentinel imagery. This joint use of census data and remote sensing produces a globally harmonized population grid suitable for cross-country and longitudinal urban analyses.

\subsection{Region selection}\label{app:regionselection}

From the CBS dataset we select 109 urban regions in the Netherlands, centered on municipalities spanning the full range of urbanization levels defined in Table~\ref{tab:urbanization_levels}. For each municipality, a rectangular bounding box is constructed that includes the municipality and part of the surrounding area. The bounding box dimensions are chosen to be an integer multiple of 32 base grid cells in both spatial directions (east–west and north–south). This ensures that the largest coarse-graining scale ($3.2 \times 3.2$ km) partitions each region into an integer number of boxes. To minimize sensitivity to the arbitrary placement of region boundaries, the coarse-graining procedure is repeated by offsetting the grid origin diagonally in both the northeast-southwest and the northwest-southeast direction in steps of 2 grid cells until a maximum of 16. These offsets shift the partitioning of the city at larger aggregation scales, and all reported results are averaged over the ensemble of the original grid together with the 32 offsets.

\begin{table}
    \centering
    \caption{Definition of urbanization levels following Statistics Netherlands.
    \label{tab:urbanization_levels} }
    \begin{tabular*}{\columnwidth}{@{\extracolsep\fill}ccc@{\extracolsep\fill}}
    \\
    \hline
    \textbf{Level} & \textbf{Description} & \textbf{addresses/km$^2$} \\
    \hline
    1 & very highly urbanized & $\geq 2.500$   \\
    2 & highly urbanized & $1.500-2.500$  \\
    3 & moderately urbanized & $1.000 - 1.500$  \\
    4 & slightly urbanized &$ 500 - 1.000$  \\
    5 & non-urbanized &  $< 500$ \\
    \hline
    \end{tabular*}
\end{table}

The selection of the 109 municipalities was guided by several criteria. First, we ensured that each of the five CBS urbanization classes was well represented. Second, municipalities whose spatial extent was too small relative to the bounding box, such that only a few boxes remained at the largest aggregation scale, were excluded. Third, municipalities with lower urbanization levels that lie directly adjacent to highly urbanized cities were removed if their bounding boxes substantially overlapped with those of larger neighboring cities. Applying these criteria resulted in a final sample of 109 urban regions. The full list of municipalities, along with their inferred exponents, regression standard errors and $R^2$ values for the fits (for the year 2023) is given in supplementary Table~\ref{tab:region_result}. 
The epoch of our data is a period that saw a large rate of merging of municipalities in the Netherlands, resulting in a narrowing of the log-normal distributions of populations per municipality. However, smaller municipalities that are not separately distinguished are in a distributional sense not very distinct from large urban regions \cite{jokic2023time}. 

For the GHS-POP dataset, we constructed a curated sample of major cities across all inhabited continents. The aim was not to obtain an exhaustive or strictly population-ranked census of global cities, but to assemble a geographically broad set of sufficiently large urban regions suitable for the same multiscale coarse-graining analysis. We sought to include at least one representative city per country, typically the principal urban center or capital, while larger countries could contribute multiple cities. The resulting sample is therefore dominated by large metropolitan areas, consistent with our focus on mature urban systems, which are expected to approach monofractal spatial organization \cite{batty1994fractal,batty2013new,murcio2015multifractal}.

For each selected city we defined a bounding box using a procedure analogous to that used for the Dutch dataset, but with dimensions chosen as multiples of $12.8$ km in order to accommodate two additional coarse-graining steps. Candidate regions were then manually inspected, and cities with substantially overlapping bounding boxes were removed to avoid double counting of urban areas. Applying these criteria resulted in a final sample of 368 cities. The full list of cities used for the analysis in this paper is displayed in the supplementary Table~\ref{tab:continent_regions}. All analysis was performed in the native World Mollweide equal-area projection (EPSG:54009) of the GHSL dataset. 

\begin{figure*} 
	\centering
	\includegraphics[width=0.9\textwidth]{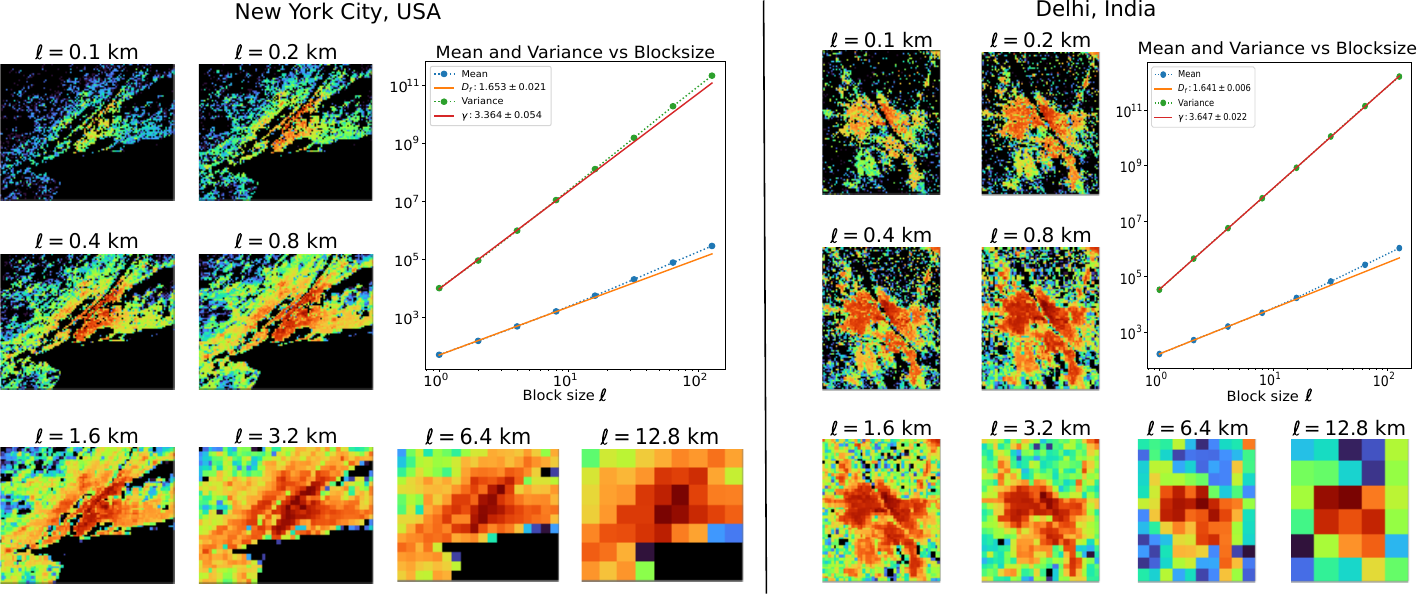} 
	\caption{\textbf{Coarse graining analysis for two cities from the GHS-POP dataset} (Left) New-York City (USA) and (Right) Delhi (India).}
	\label{fig:RG_cities}
\end{figure*}

In both datasets, regions were defined using the administrative boundaries of the corresponding city or municipality at the time of selection. The resulting bounding boxes were subsequently treated as fixed and were applied uniformly across all years of the dataset. This approach ensures that the scaling analysis is performed on the same geographic region over time, allowing for consistent longitudinal comparison as urban populations evolve. We emphasize that this longitudinal analysis compares the static multiscale spatial distribution of population across successive discrete time points, rather than computing localized temporal population growth rates.

\subsection{Coarse-graining and exponent estimation}\label{sec:coarsegraining}
Population data are aggregated on square grids of size $\ell_k=\{1,2,4,8,16,32\}\times \ell_0$, where $\ell_0=100$ m is the base grid resolution. At each scale we compute the distribution of coarse-grained population counts $N_\ell$ over non-empty boxes. A box is considered non-empty if its total population exceeds zero. In the CBS dataset this corresponds to counts $\geq 5 $ due to disclosure control. To obtain a uniform analysis, we also rounded the GHS-POP dataset to unit multiples of 5. Restricting the analysis to non-empty boxes reduces sensitivity to uninhabitable land and to the arbitrary geometry of bounding boxes. The scaling exponents $\beta$ and $\gamma$ are then estimated from a linear least squares regressions of the logarithms on the mean and variance of $N_\ell$ as functions of $\log \ell$.

Exponent estimates are obtained using the first four coarse-graining levels, corresponding to the spatial range over which the empirical scaling most closely follows a power law. Furthermore, in the small length scale limit, $\beta$ will correspond to the planar fractal dimension of populated space. On larger spatial scales, $\beta$ gradually approaches the Euclidean dimension of the plane ($\beta \to 2$) for cities not strongly constrained by uninhabitable areas such as oceans or mountain ranges, since sufficiently large boxes increasingly tile the populated region. Figure~\ref{fig:RG_cities} illustrates this behavior for two representative cities, where the slopes of both the mean and the variance increase slightly at larger $\ell$. These examples demonstrate that the scaling relations hold most accurately over an intermediate spatial range ($\ell \sim 0.1$–$2$ km), emphasizing the importance of high-resolution data for estimating the underlying fractal dimension.

\subsection{Exponent robustness and sensitivity}
\label{sec:exponent_robustness}

The scaling exponents for individual cities are estimated from a relatively small number of coarse-graining scales. Although the corresponding log--log regressions generally provide excellent fits, with $R^2$ values close to unity (Table~\ref{tab:region_result}), the limited number of fitted points raises the question of how sensitive the inferred exponents are to the precise choice of fitting scales. We therefore perform an additional robustness analysis for the Netherlands in 2024.

The analysis in the main text focuses on the small-$\ell$ scaling regime. At sufficiently large $\ell$, the spatial scaling of the mean population crosses over toward the Euclidean limit $\beta=2$, so extending the fitting range to progressively larger coarse-graining scales does not necessarily provide a better estimate of the small-scale exponent. Instead, to obtain a denser sampling of the relevant range of scales, we repeat the aggregation using linearly spaced box sizes,
\begin{equation}
    \ell_k = k\ell_0,
    \qquad
    k=1,2,\ldots,15,
\end{equation}
rather than the exponentially spaced sequence $\ell_k=\{1,2,4,8,\ldots\}\ell_0$ used in the main analysis. We then construct a sequence of nested fitting windows,
\begin{equation}
    \ell/\ell_0 = \{1,2,\ldots,k\},
    \qquad
    k=5,\ldots,15,
\end{equation}
and re-estimate $\beta$ and $\gamma$ by linear regression in log--log space for each window. Importantly, these fits do not reproduce the main-text fitting procedure: the latter uses the four logarithmically spaced scales $\{1,2,4,8\}\ell_0$. The comparison therefore probes sensitivity both to the extent of the fitting range and to the sampling of scales within that range.

\begin{figure}
    \centering
    \includegraphics[width=\columnwidth]{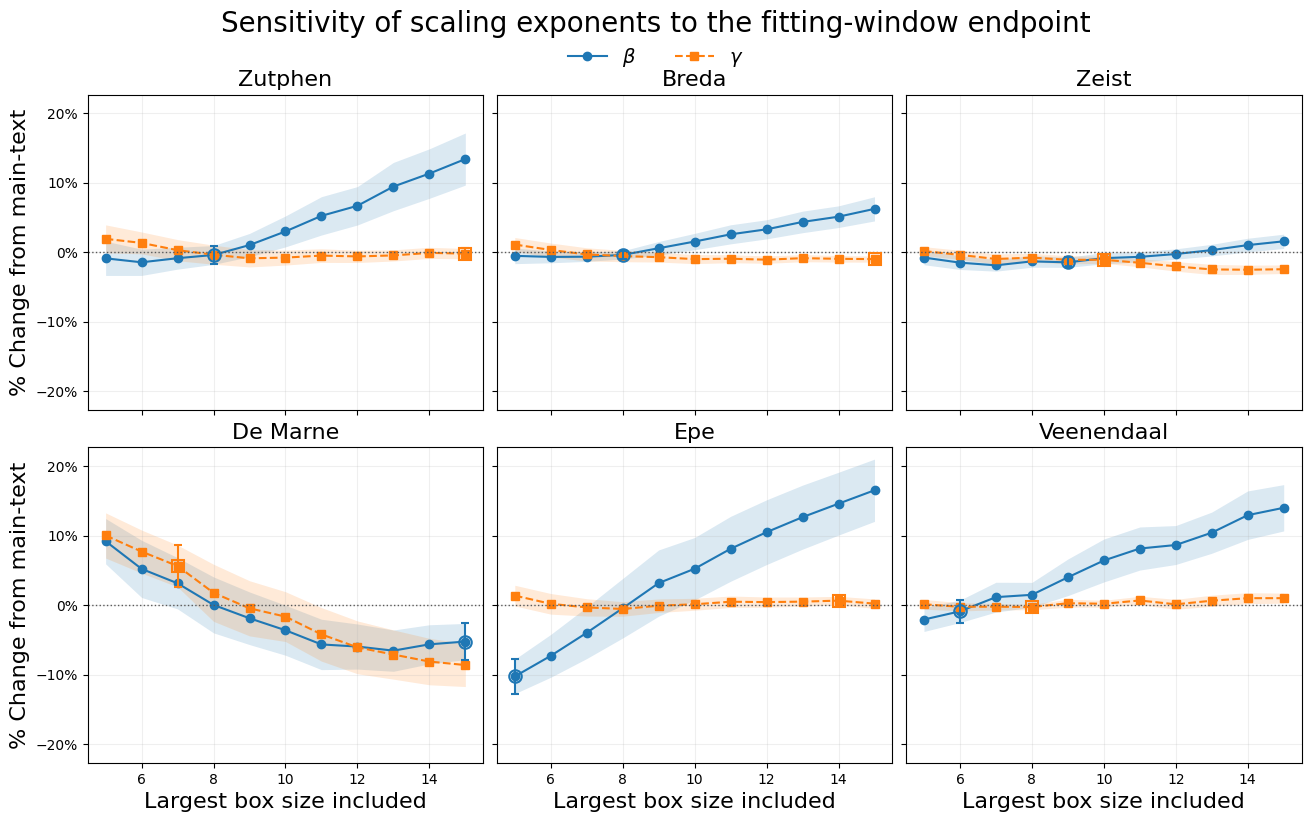}
    \caption{
    Relative change in the estimated exponents $\beta$ (blue circles) and $\gamma$ (orange squares) as the endpoint of the nested fitting window is increased. Changes are measured relative to the exponents reported in the main text. The six cities are selected to represent small ($\beta<1$; left column), intermediate ($1\leq\beta\leq1.3$; middle column), and large ($\beta>1.3$; right column) values of $\beta$. Within each range, the upper and lower panels show examples with comparatively weak and strong fitting-window sensitivity, respectively. Shaded regions indicate the regression uncertainty (standard deviation) of the exponent estimates and bold markers indicate the best fit according to the smallest regression uncertainty.
    }
    \label{fig:rep_city_robustness}
\end{figure}

Figure~\ref{fig:rep_city_robustness} illustrates the resulting variation for six representative cities. The examples were deliberately selected to span both the range of $\beta$ values and the observed range of fitting-window sensitivities, and therefore include some of the more pronounced deviations in the dataset. For most fitting windows, changes remain at the level of a few percent, although the most sensitive examples can reach approximately $10$--$20\%$ when the fitting range is extended to the largest scales considered.

The changes are predominantly smooth and systematic rather than irregular. For several cities, $\beta$ increases as larger box sizes are included, whereas $\gamma$ varies less and often decreases slightly. There are also cities, such as De Marne, for which both exponents decrease as the fitting range is extended. These trends indicate that the fitting-window dependence is not primarily generated by statistical fluctuations in the regressions. Rather, it is consistent with a gradual crossover in the effective scaling behavior as progressively larger spatial scales are included. This observation also motivates retaining a restricted small-scale fitting range when interpreting $\beta$ as a fractal dimension.

\begin{figure}
    \centering
    \includegraphics[width=\columnwidth]{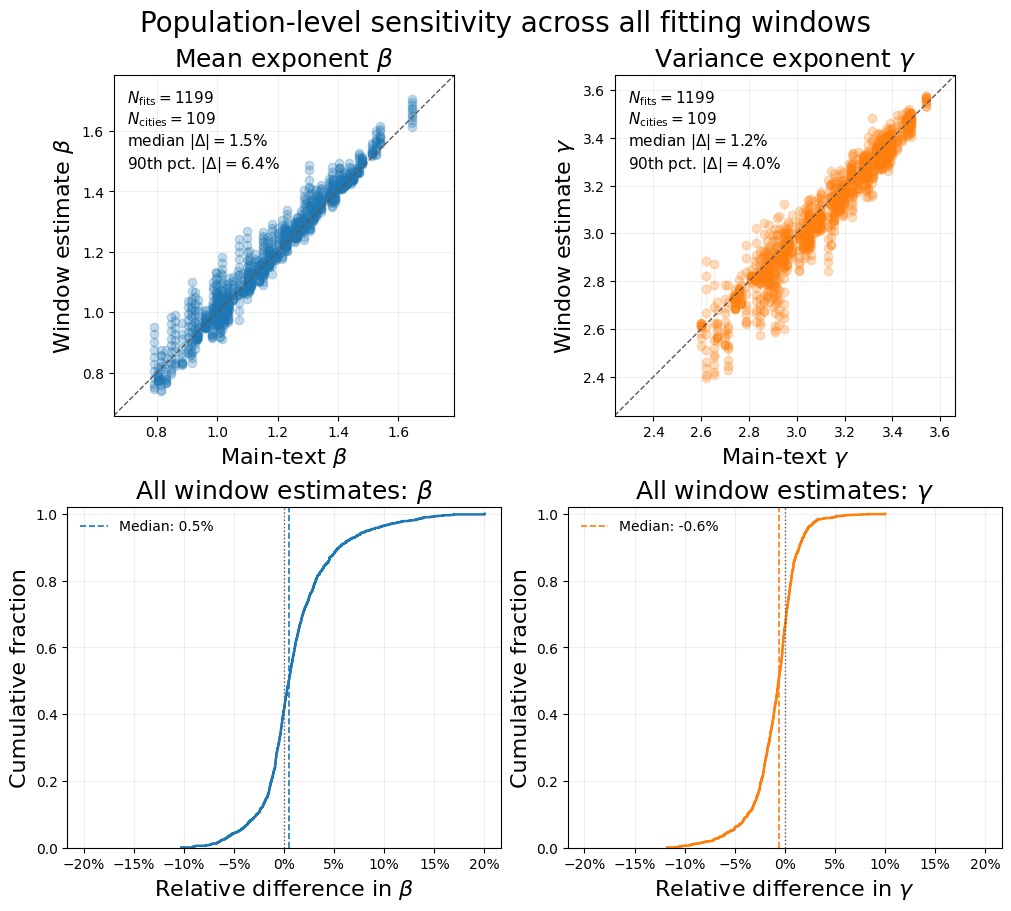}
    \caption{
    Population-level sensitivity of the scaling exponents to the fitting-window endpoint. The upper panels compare all exponent estimates obtained from the 11 linearly sampled fitting windows with the corresponding main-text estimates for all 109 cities; dashed lines indicate equality. The lower panels show the empirical cumulative distribution functions (ECDF) of the relative differences. Each city contributes the same number of fitting windows.
    }
    \label{fig:population_sensitivity}
\end{figure}

The population-level results are shown in Fig.~\ref{fig:population_sensitivity}. Across the 109 cities and 11 fitting windows, this procedure yields $1199$ estimates for each exponent. Despite the different choice of fitting scales, the estimates remain concentrated close to the values reported in the main text. The median absolute relative difference is only $1.5\%$ for $\beta$ and $1.2\%$ for $\gamma$. Moreover, $90\%$ of the estimates differ from the main-text values by less than $6.4\%$ for $\beta$ and $4.0\%$ for $\gamma$.

For the densest $k=15$ fits, the median $R^2$ is $0.997$ for $\beta$ and $0.999$ for $\gamma$, with $90\%$ of cities having $R^2$ values above $0.985$ and $0.994$, respectively. These values should, however, be interpreted primarily as descriptive measures of the linearity of the log--log scaling curves. The observations across $\ell$ are obtained by nested coarse-graining of the same underlying spatial field and are therefore not statistically independent, so the near-unity $R^2$ values do not carry the same inferential meaning as in an ordinary regression with independent observations.

The empirical cumulative distribution functions (ECDF) in the lower panels of Fig.~\ref{fig:population_sensitivity} show that the signed deviations are also centered close to zero. The median relative change is $+0.5\%$ for $\beta$ and $-0.6\%$ for $\gamma$, indicating that there is no substantial systematic shift relative to the main-text estimates. The somewhat broader distribution for $\beta$ is consistent with the stronger fitting-window dependence visible for the mean exponent in Fig.~\ref{fig:rep_city_robustness}.

The $1199$ estimates should not be interpreted as statistically independent observations, because fitting windows belonging to the same city overlap strongly. The ECDFs therefore characterize the sensitivity of the estimates across fitting specifications rather than the sampling distribution of $1199$ independent measurements. Since every city contributes exactly 11 fitting windows, however, this representation weights all cities equally and provides a direct summary of the dependence on the fitting range.

\begin{figure}
    \centering
    \includegraphics[width=\columnwidth]{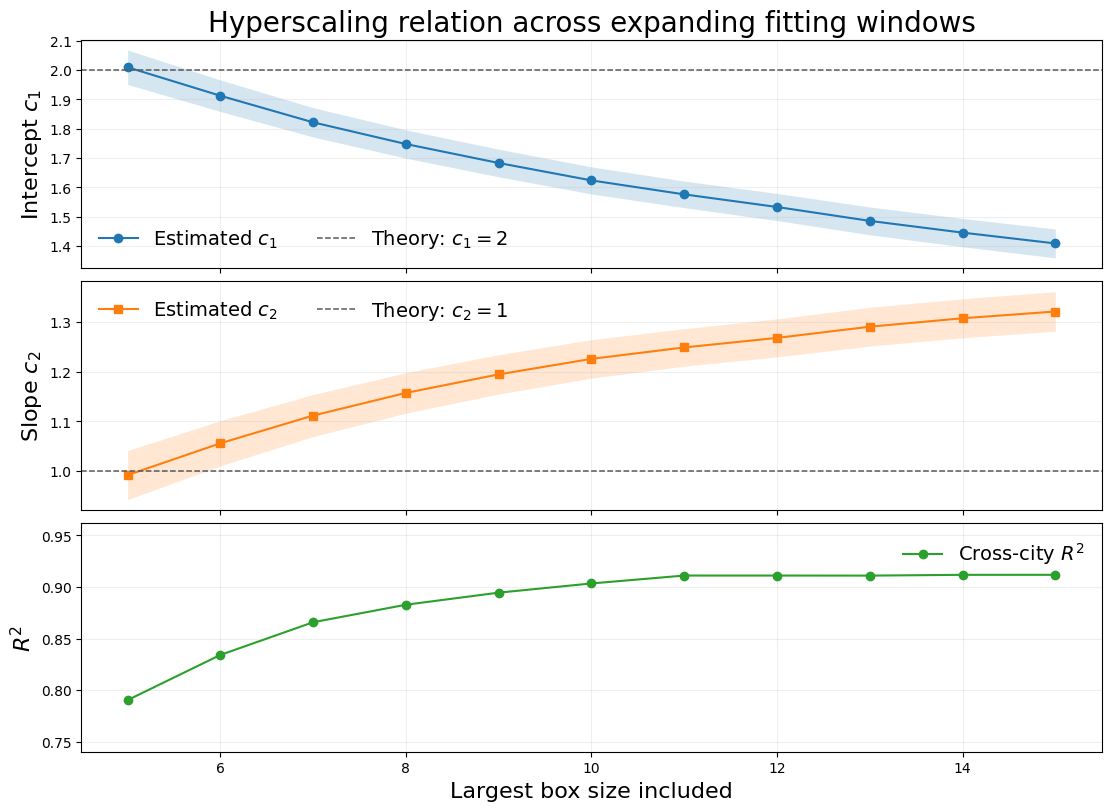}
    \caption{
    Robustness of the cross-city hyperscaling relation to the fitting-window endpoint. For each endpoint $k$, the exponents $\beta_i(k)$ and $\gamma_i(k)$ are re-estimated for all 109 cities and the relation
    $\gamma=c_1+c_2\beta$
    is fitted across cities. The upper and middle panels show the fitted intercept $c_1$ and slope $c_2$, with shaded regression uncertainties; dashed lines indicate the theoretical values $c_1=2$ and $c_2=1$. The lower panel shows the corresponding coefficient of determination.
    }
    \label{fig:hyperscaling_robustness_test}
\end{figure}

Finally, we test whether the cross-city hyperscaling relation itself depends on the fitting range. For each endpoint $k$, we fit
\begin{equation}
    \gamma_i(k)=c_1(k)+c_2(k)\beta_i(k)
\end{equation}
across the same 109 cities. The results are shown in Fig.~\ref{fig:hyperscaling_robustness_test}. The exponent relation remains strong over the complete range of fitting windows. Indeed, its $R^2$ increases from approximately $0.79$ for the smallest window to above $0.9$ for the largest windows. At the same time, the fitted coefficients vary systematically with the fitting range: the intercept $c_1$ decreases, while the slope $c_2$ increases.

A particularly striking result is that, for the smallest linearly sampled fitting window, the coefficients are very close to the theoretical prediction,
\begin{equation}
    c_1=2,
    \qquad
    c_2=1.
\end{equation}
This agreement is not imposed by the main-text fitting procedure. The smallest robustness window uses the linearly spaced scales $\{1,2,3,4,5\}\ell_0$, whereas the main-text exponents are obtained from the logarithmically spaced scales $\{1,2,4,8\}\ell_0$. The recovery of approximately the same theoretical hyperscaling relation under these distinct choices of scales therefore suggests that the relation is characteristic of the fine-scale regime itself rather than an artifact of a particular discretization of the fitting range.

As progressively larger scales are included, the cross-city relation remains well defined and even becomes statistically tighter, but its coefficients continuously depart from $c_1=2$ and $c_2=1$. We interpret this as evidence for a scale-dependent crossover: the existence of a strong relation between $\beta$ and $\gamma$ is robust, whereas its precise coefficients characterize the spatial regime over which the individual exponents are measured. Together, these tests show that the main empirical conclusions are not sensitive to modest variations of the fitting procedure, while also emphasizing that the exponents should be interpreted as effective small-scale scaling exponents rather than scale-independent quantities.

\subsection{Hyperscaling relations over time}
\label{app:temp_hyperscaling}
\begin{figure*}
\centering
\includegraphics[width=\textwidth]{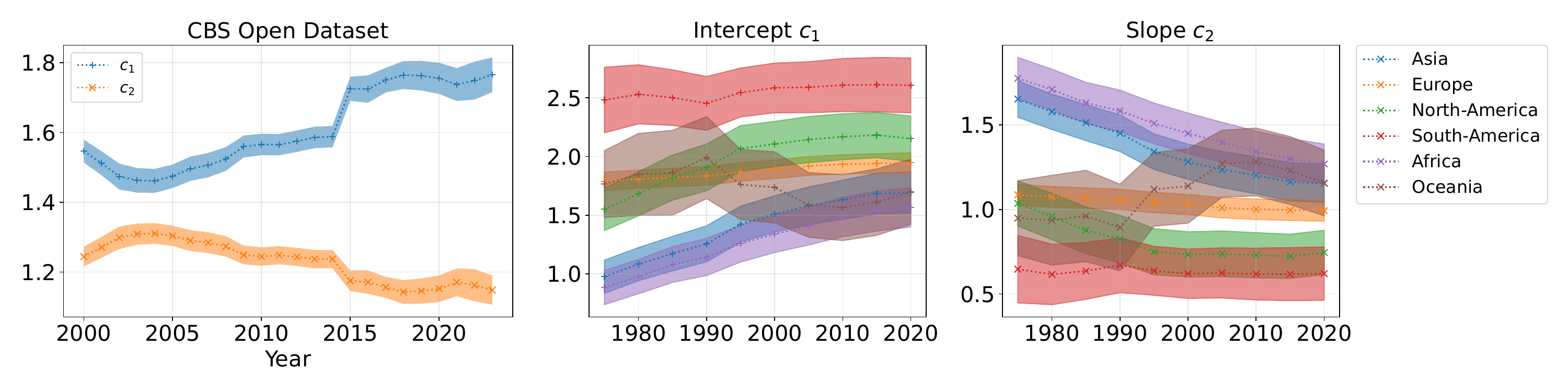}
\caption{\label{fig:hyperscale_parameters} (Left) Hyperscaling parameters $c_1$ and $c_2$ for the equation $\gamma = c_1 + c_2 \beta$ for the subset of Dutch cities chosen for this study in different years. (Middle \& Right) Hyperscale parameters $c_1$ and $c_2$ for cities grouped by their continents. The band indicates standard deviation from the linear regression of the inferred parameters.}
\end{figure*}

The scaling analysis described above was applied to CBS population data for each year from 2000 to 2023 \cite{cbs2024dataset}. For each year we estimate the relationship between the fractal dimension $\beta$ and the variance exponent $\gamma$ using a linear regression of the form
\begin{equation}\label{eq:hyperscale}
\gamma = c_1 + c_2 \beta .
\end{equation}
The resulting coefficients $c_1$ and $c_2$ are shown in the left panel of Fig.~\ref{fig:hyperscale_parameters} for the CBS dataset. Their temporal variability exceeds the regression uncertainties (indicated by the shaded regions), suggesting that the hyperscaling relation is sensitive to temporal changes associated with urban development and population dynamics. Over the observation period we observe a gradual increase in $c_1$ and a corresponding decrease in $c_2$.

A noticeable discontinuity occurs between the years 2014 and 2015. This reflects a change in the CBS data release format: data for 2000–2014 were distributed as a single dataset processed using a consistent methodology, whereas from 2015 onward the data are released annually with substantially more information per grid cell. Differences in statistical disclosure procedures may therefore affect pixels with small population counts. Indeed, the post-2015 datasets contain systematically more cells with the minimum reported population value (five inhabitants) than the earlier data. This observation further highlights the sensitivity of scaling analyses to the treatment of low-count cells in fine-resolution population grids.

The middle and right panels of Fig.~\ref{fig:hyperscale_parameters} show the inferred coefficients $c_1$ and $c_2$ when subdividing the GHS-POP dataset by continent. We see that while European cities already show a hyperscaling relation close to $\gamma = 2+\beta$ for all years in the sample, cities from Africa and Asia show a remarkable change of $c_1 \to 2$ and $c_2 \to 1$ over the years. This coincides with a period of rapid urbanization for major cities in these continents, which could be responsible for the transition to monofractal structure. Interestingly, cities on the American continent seem to overshoot the relation $\gamma = 2 + \beta$ predicted by monofractality and strong spatial correlations. Indeed both South and North American cities seem to show $c_1>2$ and $c_2 < 1$, which is particularly pronounced for South-America. In this case, we have even removed a outlier in the dataset; the city of Lima, Peru, as this city shows an extremely low $\gamma$ for its fractal dimension. 


Several factors may contribute to these deviations. First, geographical constraints can strongly influence urban morphology; for example, the growth of Lima is restricted by both the Pacific Ocean and the Andes mountain range. Second, many American cities exhibit extensive suburban regions characterized by relatively homogeneous low-density development, which can increase $\beta$ while reducing fluctuation intensity, thereby shifting the inferred coefficients. Finally, uncertainties in population estimates of rapidly developing or informally settled urban areas may introduce biases in the gridded population products. Further work would be required to disentangle these effects.

\section{Mean-field theory}\label{app:meanfield}
We derive here the mean--variance relation expected under a mean-field approximation in which fractal geometry enters through an average occupation probability, while fluctuations are described by finite and homogeneous variance and covariance. This leads to a quadratic mean--variance relation of Taylor type, at odds with the empirically observed linear relation between $\beta$ and $\gamma$.

Let $N_{k+1}$ denote the population in a coarse-grained block at level $k+1$, obtained by aggregating $b\times b$ subcells from level $k$:
\begin{equation}
N_{k+1}=\sum_{i=1}^{b^2}\xi_i X_i .
\end{equation}
Here $X_i$ is the population of subcell $i$, while $\xi_i\in\{0,1\}$ indicates whether that subcell belongs to the populated support. We assume spatially homogeneous mean-field statistics at level $k$,
\begin{align}
& \mathbb{E}[X_i] =\mu_k, \qquad \mathrm{Var}(X_i) =v_k, \nonumber \\
& \mathrm{Cov}(X_i,X_j) =c_k \quad (i\neq j),
\end{align}
and encode the fractal geometry through the average occupation probability
\begin{equation}
\mathbb{E}[\xi_i]=p_k,\qquad b^2p_k=b^{d_f},
\end{equation}
so that the expected number of populated subcells in each coarse-graining block scales as $b^{d_f}$.

Under this approximation, the mean obeys
\begin{equation}
\mu_{k+1}=\mathbb{E}[N_{k+1}]
= b^2p_k\,\mu_k
= b^{d_f}\mu_k.
\end{equation}
Similarly, the variance satisfies
\begin{align}
v_{k+1}
= \mathrm{Var}(N_{k+1})
& = b^2p_k\,v_k + b^2p_k(b^2p_k-1)c_k \nonumber \\ 
& = b^{d_f}v_k + b^{d_f}(b^{d_f}-1)c_k,
\end{align}
while the covariance between distinct coarse-grained blocks obeys
\begin{equation}
c_{k+1}=b^{2d_f}c_k.
\end{equation}

These results form a set of recursive relation for the mean, variance and co-variance at each coarse-graining level in terms of that at a lower level. Their solution is:
\begin{align}\label{MFsol}
\mu_k & =\mu_0(b^k)^{d_f}, \nonumber \\
v_k &=(v_0-c_0)(b^k)^{d_f}+c_0(b^k)^{2d_f}, \nonumber \\
c_k & =c_0(b^k)^{2d_f}.
\end{align}
Here $\mu_0, v_0$ and $c_0$ are the mean, variance and covariance at level 0. 
Defining the coarse-graining length scale as $\ell=b^k$, we obtain
\begin{equation}
\mu(\ell)\propto \ell^{d_f},
\qquad
v(\ell)=a\,\ell^{d_f}+b\,\ell^{2d_f},
\end{equation}
with $a=v_0-c_0$ and $b=c_0$. Hence, within mean field, the variance is not a pure power law but
a sum of two contributions. Depending on the relative magnitude of variance and covariance at the finest scale, an effective exponent $\gamma$ inferred from a single power-law fit may therefore lie between $d_f$ and $2d_f$.

Eliminating the latent scale variable $\ell$ from Eq.~\eqref{MFsol} yields a quadratic mean–variance relationship:
\begin{equation}\label{Taylors}
v_k =
\frac{v_0-c_0}{\mu_0}\mu_k +
\frac{c_0}{\mu_0^2}\mu_k^2,
\qquad
c_k = \frac{c_0}{\mu_0^2}\mu_k^2 .
\end{equation}
Thus Taylor’s law emerges naturally from spatial aggregation under the mean-field assumption. Because the derivation does not depend on the specific nature of the data, the same mechanism applies to other spatial systems where Taylor-type scaling is observed (e.g., ecological or epidemiological data). 

However, the mean-field analysis also indicates that $\gamma \in [d_f, 2d_f]$, whereas we observed empirically that evolving urban regions tend towards $\gamma \sim 2 + d_f > 2d_f$. This discrepancy invalidates the mean-field assumption, indicating that spatial correlations play a crucial role in the observed fluctuation scaling. This motivates the correlation-aware variance decomposition derived in the next section.

\section{Spatial correlations and the monofractal limit}
\label{app:correlations}
The mean-field analysis above shows that when spatial units contribute independently with finite (co)variance, the coarse-grained variance generically takes the polynomial form
\begin{equation}
\mathrm{Var}(N_\ell) = a\,\ell^{\beta} + b\,\ell^{2\beta}.
\end{equation}
Such a quadratic mean–variance relation cannot reproduce the empirically observed linear interdependence between the exponents $\beta$ and $\gamma$. This indicates that spatial correlations extending beyond the mean-field approximation must play a central role. In particular, the empirical data show a systematic approach toward the relation $\gamma = 2 + \beta$ in large urban systems. We now derive how the variance scales when spatial correlations are explicitly taken into account.

Let $X_i$ denote the population in fine-grid cell $i$ of side length $\ell_0$. For a square box $A_\ell$ of side $\ell$, the coarse-grained population is
\begin{equation}
N_\ell = \sum_{i \in A_\ell} X_i .
\end{equation}
The variance of $N_\ell$ can be decomposed into independent and correlated contributions,
\begin{align}\label{Var_decomposition}
\mathrm{Var}(N_\ell)
& = 
\sum_{i \in A_\ell} \mathrm{Var}(X_i)
+
\sum_{\substack{i,j \in A_\ell \\ i \neq j}}
\mathrm{Cov}(X_i,X_j) \nonumber \\
& =: V_1(\ell) + V_2(\ell).
\end{align}
If the local variance $\mathrm{Var}(X_i)$ remains finite and does not scale with $\ell$, the first term simply counts the number of summands. Since a square region of side $\ell$ contains $n(\ell) \propto \ell^2$ cells,
\begin{equation}
V_1(\ell) \propto n(\ell) \propto \ell^2 .
\end{equation}
Importantly, this contribution depends only on the geometric number of summands and not on their mean occupancy. Even if the mean population scales as $\langle N_\ell \rangle \sim \ell^\beta$, empty cells contribute $X_i = 0$ and therefore add zero to the variance while still being counted among the summands. Consequently,
\begin{equation}
V_1(\ell) \propto \ell^2,
\end{equation}
which by itself would imply the fluctuation exponent $\gamma = 2$. 

The second term in \eqref{Var_decomposition} captures spatial dependence between population values in nearby cells. Assuming approximate statistical homogeneity, we introduce the covariance function
\begin{equation}
C(r) = \mathrm{Cov}(X(\mathbf{x}),X(\mathbf{y})),
\qquad
r = |\mathbf{x}-\mathbf{y}|.
\end{equation}
For sufficiently large $\ell$, the discrete double sum can be approximated by a double integral,
\begin{equation}
V_2(\ell)
\approx 
\int_{A_\ell} \!\! d^2 x
\int_{A_\ell} \!\! d^2 y \;
C(|\mathbf{x}-\mathbf{y}|).
\end{equation}
Performing the change of variables $\mathbf{u} = \mathbf{y}-\mathbf{x}$,
$\mathbf{v}=\mathbf{x}$ and neglecting boundary corrections at large $\ell$,
one obtains
\begin{equation}
V_2(\ell)
\approx 
|A_\ell| 
\int_{A_\ell} C(|\mathbf{u}|)\, d^2 u
= \ell^2
\int_{A_\ell} C(r)\, d^2 r.
\end{equation}

Suppose now that the spatial correlations decay as a power law,
\begin{equation}
C(r) \sim r^{-\eta},
\qquad
0 \le \eta < 2.
\end{equation}
Approximating the integration domain by a disk of radius $\ell$ yields
\begin{equation}
\int_{A_\ell} C(r)\, d^2r
\sim
\int_0^\ell r^{-\eta} 2\pi r\, dr
\propto
\ell^{2-\eta}.
\end{equation}
Consequently,
\begin{equation}
V_2(\ell) \propto \ell^2 \ell^{2-\eta}
= \ell^{4-\eta}.
\end{equation}

It is convenient to define the \emph{correlation dimension}
\begin{equation}
D_c = 2 - \eta,
\end{equation}
so that the covariance contribution scales as
\begin{equation}
V_2(\ell) \propto \ell^{2+D_c}.
\end{equation}

Combining the independent and correlated contributions of \eqref{Var_decomposition} gives
\begin{equation}
\mathrm{Var}(N_\ell)
=
A\,\ell^2
+
B\,\ell^{2+D_c},
\label{eq:var_scaling_correlated}
\end{equation}
where $A,B$ are constants quantifying the relative strength of both contributions. Using Eq.~\eqref{eq:var_scaling_correlated}, the effective variance exponent can be written as
\begin{align}
\gamma(\ell)
&=
\frac{d \log \mathrm{Var}(N_\ell)}{d \log \ell}
=
\frac{2A\ell^2 + (2+D_c)B\ell^{2+D_c}}{A\ell^2 + B\ell^{2+D_c}} \nonumber \\
&=
2 + \frac{D_c\,R(\ell)}{1+R(\ell)},
\qquad
R(\ell)=\frac{B\,\ell^{D_c}}{A}.
\end{align}
It follows immediately that
\begin{equation}
2 \le \gamma(\ell) \le 2 + D_c,
\label{eq:gamma_bound}
\end{equation}
with the lower bound attained when the independent contribution dominates and the upper bound attained in the correlation-dominated regime.

Let $\beta$ denote the mass-fractal dimension defined by
\begin{equation}
\langle N_\ell \rangle \sim \ell^\beta .
\end{equation}
In general multifractal systems the correlation dimension $D_c$ need not equal $\beta$. However, when the spatial distribution approaches a monofractal regime, the generalized dimensions converge:
\begin{equation}
D_c = D_2 = \beta .
\end{equation}
An approach to monofractality is empirically plausible for large mature cities \cite{batty1994fractal,murcio2015multifractal}. In that case, $D_c=\beta$, and Eq.~\eqref{eq:gamma_bound} becomes
\begin{equation}
2 \le \gamma(\ell) \le 2 + \beta.
\end{equation}
Hence the observed drift toward $\gamma \simeq 2+\beta$ can be interpreted as a crossover from an independent-fluctuation regime to a correlation-dominated monofractal regime.
As cities evolve toward monofractal organization, the correlation dimension and the mass dimension coincide, and the scaling of fluctuations becomes geometrically constrained by the fractal support of the population field.

In summary, the variance decomposition in Eq.~\eqref{Var_decomposition} shows that the observed fluctuation scaling results from the competition between an independent contribution, which enforces $\gamma = 2$, and a covariance contribution, which increases the effective exponent toward $\gamma = 2 + D_c$. In this sense, spatial correlations do not merely renormalize the prefactor of the variance, but change its scaling structure across length scales. When the population field further approaches a monofractal regime such that $D_c \simeq \beta$, the fluctuation exponent is driven toward the asymptotic hyperscaling form $\gamma \simeq 2+\beta$.\footnote{When generalizing the argument to arbitrary spatial dimension, the limiting relation becomes $\gamma = d + D_c$, which further motivates the use of the phrase `hyperscaling relation' as a relation between exponents explicitly involving the spatial dimension $d$.} This provides a natural explanation for the empirical drift of the fitted $\beta$--$\gamma$ relation observed in the data and shows why any successful mechanistic model of urban growth must reproduce not only the fractal scaling of the mean population, but also the correlation structure governing its fluctuations.

\section{Urban Growth Models}\label{app:models}

The analytical results above show that the observed hyperscaling relation requires spatial correlations, and that in the monofractal limit these correlations constrain the asymptotic form toward $\gamma \simeq 2+\beta$. To illustrate these ideas in simple generative settings, we briefly examine two stylized urban growth models. The goal of this section is not to obtain a quantitative fit to the empirical $\beta$--$\gamma$ relation, but rather to show that spatially correlated growth processes naturally generate qualitatively similar behavior and to highlight what present models still fail to capture. In particular, both models below reproduce an interdependence between $\beta$ and $\gamma$, but their precise scaling curves remain sensitive to simplifying assumptions such as binary occupancy and uniform cell capacity. We therefore view the measured hyperscaling relation as an additional empirical constraint that more realistic urban growth models should satisfy.

\begin{figure*}
    \centering
    \includegraphics[width=\textwidth]{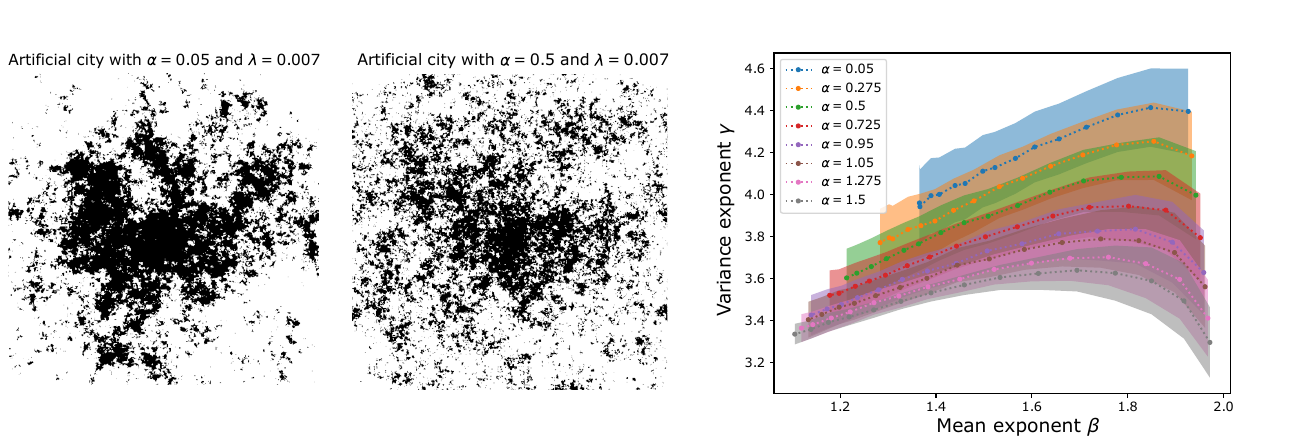}
    \caption{\textbf{Relation between scaling exponents from the correlated percolation model.}
    (Left and middle) Two sample artificial cities generated by the correlated percolation model of \cite{makse1998modeling} with $\alpha=0.05$ and $\alpha=0.5$. Both examples have $\lambda=0.007$ and lattice size $512\times512$.
    (Right) Relation between the mean exponent $\beta$ and the variance exponent $\gamma$ in the correlated percolation model. Each color corresponds to a different value of the correlation parameter $\alpha$ (smaller $\alpha$ corresponds to stronger correlations). Along each curve, the data points correspond to decreasing $\lambda$ from left to right, between $0.015$ and $0.001$, so that $\lambda$ controls the density and extent of the artificial city. Each point represents an average over 100 independently generated realizations with identical parameter values; shaded regions indicate the corresponding standard deviation.}
    \label{fig:rg_percolation_model}
\end{figure*}

\subsection{Correlated percolation model}
We first consider the correlated percolation model for urban growth introduced in \cite{makse1998modeling}. We generate artificial urban patterns using this model and apply the same multiscale aggregation pipeline as used for the empirical data. The model is defined on a two-dimensional lattice of size $(L,L)$, where each cell is either occupied or empty. In ordinary percolation, cells are occupied independently. In the correlated version, the probability of occupation is spatially dependent: cells near already occupied sites are more likely to become occupied, mimicking the tendency of urban development to cluster around existing built-up areas.

Full details on the model are available in \cite{makse1998modeling}. For the purposes of the present analysis, the model is controlled by two parameters, $\alpha$ and $\lambda$. The parameter $\alpha$ controls the strength of spatial correlations through a covariance kernel of the form
\begin{equation}
C(\ell)=\langle u(r)u(r')\rangle=(1+\ell^2)^{-\alpha/2},
\qquad
\ell = |r-r'|.
\end{equation}
Smaller $\alpha$ corresponds to stronger correlations, while $\alpha \geq 2$ approaches the uncorrelated percolation limit. The second parameter, $\lambda$, controls the mean density profile of the generated city through an exponential radial decay,
\begin{equation}
p(r)\sim e^{-\lambda r},
\end{equation}
with $r$ the distance to the center. This produces patterns that are denser in the core and progressively sparser toward the periphery.

As demonstrated in Fig.~\ref{fig:rg_percolation_model}, over the range of fractal dimensions relevant for the empirical data, approximately $1 \lesssim \beta \lesssim 1.6$, the model exhibits a clear increase of $\gamma$ with $\beta$, qualitatively similar to the urban data. More importantly, the family of curves is systematically organized by the correlation parameter $\alpha$: changing the correlation strength changes both the slope and the height of the $\beta$--$\gamma$ relation. In this sense, the model provides an explicit numerical illustration of the mechanism discussed in the previous section, namely that the fluctuation exponent is controlled not only by the geometry of the occupied support, but also by the spatial correlation structure.

At the same time, the model also makes clear why matching the empirical relation quantitatively is nontrivial. In particular, the curves bend downward as $\beta$ approaches $2$. This is not a feature of the data, but rather an artifact of the model’s binary occupancy: each cell can host at most one unit, so once the pattern becomes dense the lattice is nearly uniformly filled and the variance is suppressed. The correlated percolation model therefore supports the role of spatial correlations in shaping the hyperscaling relation, but it does not reproduce the empirical asymptotic form quantitatively. In particular, there is no reason to expect the correlation dimension in this model to coincide with the mass-fractal dimension, so the monofractal limit discussed above is not realized here.

\subsection{Dynamical urban growth model}
We next consider a simple nonequilibrium growth model defined on a two-dimensional lattice whose sites may be occupied ($A$) or empty ($\emptyset$). The model is intended as a stylized dynamical analogue of urban development, in which local occupation is promoted by nearby occupation, while random entry and exit events remain possible. The resulting dynamics generates spatially correlated, fractal-like patterns over part of parameter space.

The model can be written as a reaction system on nearest-neighbor lattice sites $i,j$,
\begin{subequations}\label{CRN_rules}
\begin{align}
A_i &\xrightarrow{\mu} \emptyset_i, \\
\emptyset_i &\xrightarrow{b} A_i, \\
A_i + \emptyset_j &\xrightarrow{rb} A_i + A_j, \\
A_i + \emptyset_j &\xrightarrow{r\mu} \emptyset_i + \emptyset_j .
\end{align}
\end{subequations}
Here $b$ is a spontaneous occupation rate, $\mu$ is a vacancy rate, and $r$ controls the strength of nearest-neighbor interactions. When $r$ is large, local interactions dominate over random updates; when $r$ is small, the system becomes more weakly correlated. By rescaling time we may set $\mu=1$ without loss of generality.

The nearest neighbor rules in \eqref{CRN_rules} are those of the voter model, where neighbors can align their states. Hence this dynamical system is a variant of voter model with asymmetric interactions and random noise (which disappears in the large $r$ limit). In this parameterization, $b$ primarily controls the overall density of the lattice, and therefore the effective fractal dimension of the generated pattern, while $r$ controls the strength of spatial correlations and hence the slope of the resulting $\beta$--$\gamma$ relation. When $b=\mu = 1$ the system lies near the critical behavior of the two-dimensional voter-model universality class \cite{dornic2001critical}.

\begin{figure*}
\centering
\includegraphics[width=\textwidth]{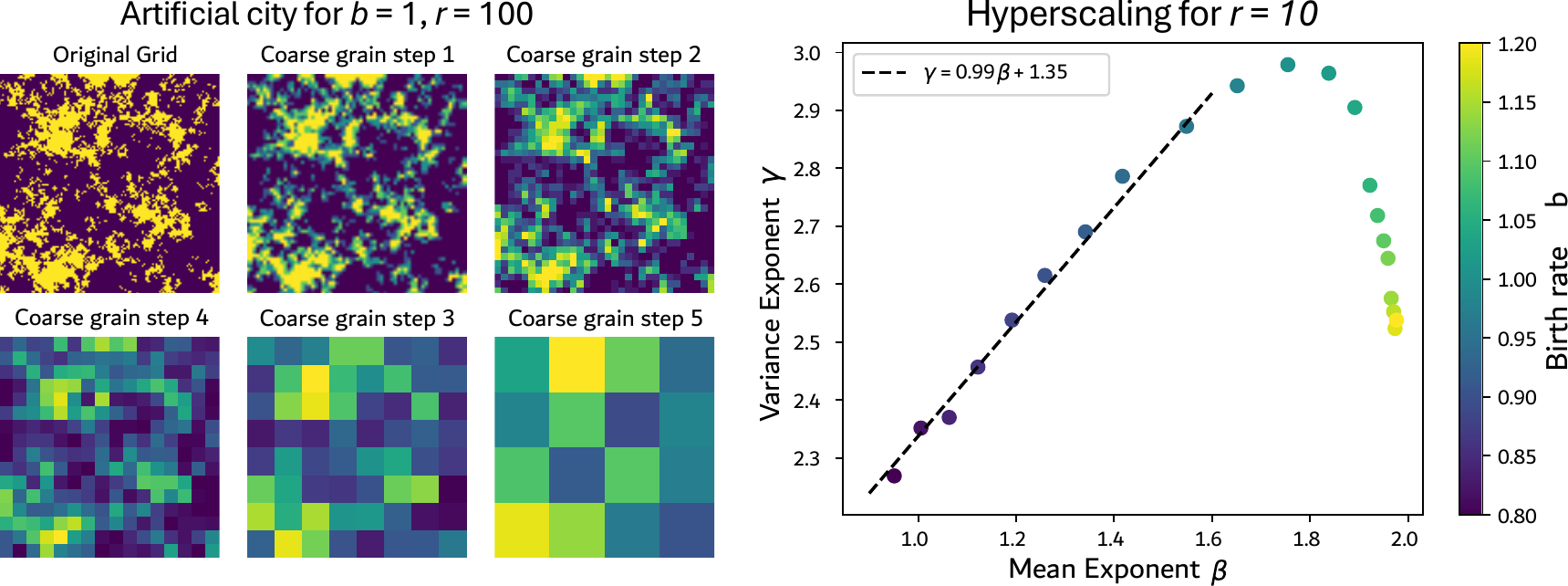}
\caption{\textbf{Hyperscaling in the dynamical urban growth model.}
(Left) Example of a fractal-like lattice generated by the dynamical model and its successive coarse-graining steps.
(Right) Relation between the exponents $\beta$ and $\gamma$ obtained from Monte Carlo samples of the model. Increasing the birth rate $b$ yields denser regions and larger $\beta$. The largest $\gamma$ values are obtained close to the critical point $b=1$, after which $\gamma$ decreases again as $\beta \rightarrow 2$ because of the model’s binary occupancy constraint.}
\label{fig:rg_fractal_model}
\end{figure*}

Figure~\ref{fig:rg_fractal_model} shows a representative configuration together with the resulting $\beta$--$\gamma$ relation obtained from Monte Carlo sampling of the model. As in the correlated percolation case, the model reproduces a qualitative increase of $\gamma$ with $\beta$ over the range most relevant to the data, followed by a drop as $\beta\rightarrow2$. Again, this downturn is an artifact of the imposed binary occupancy and uniform maximal cell capacity: once the lattice becomes nearly filled, fluctuations are artificially suppressed.

\begin{figure}
    \centering
    \includegraphics[width=\linewidth]{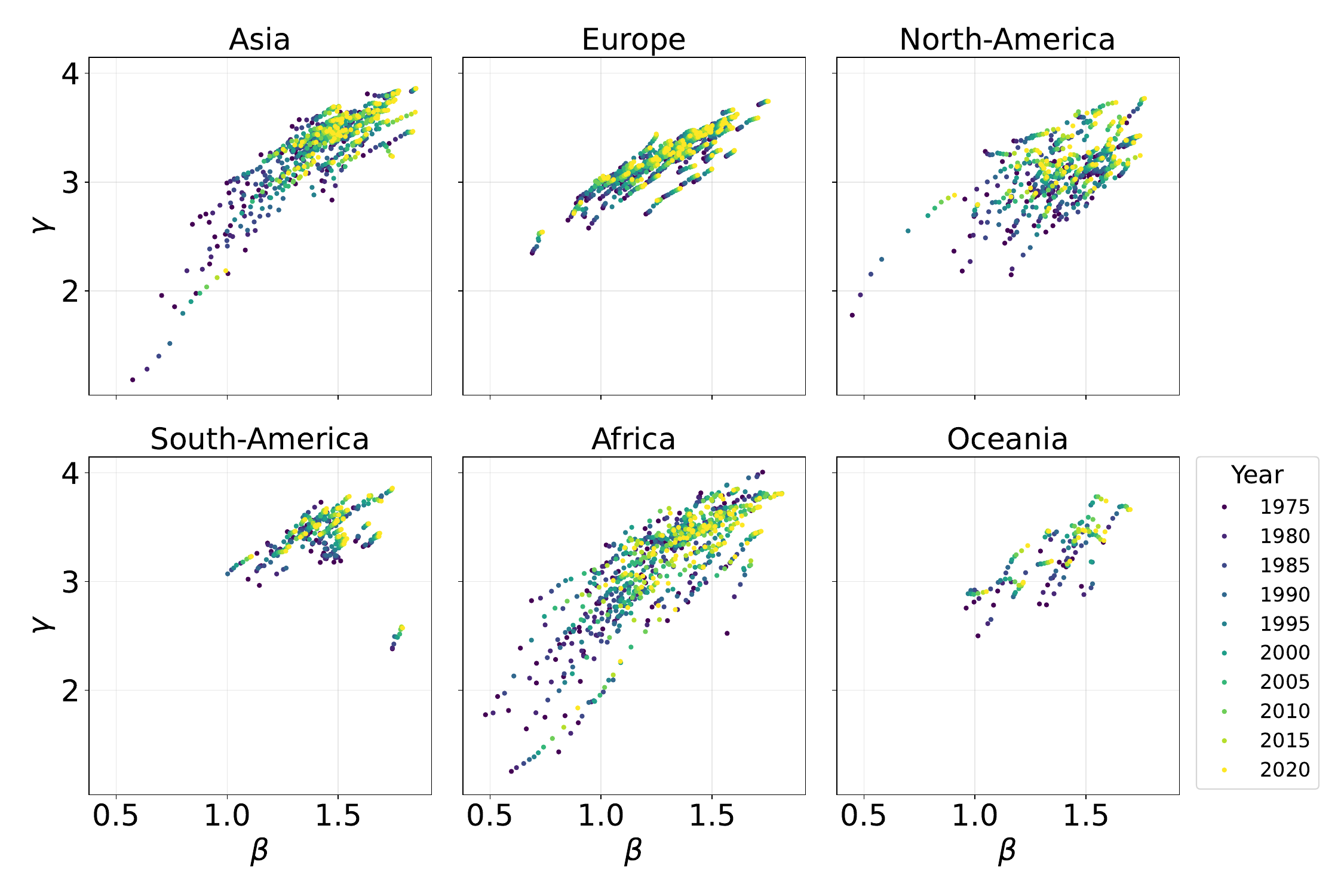}
    \caption{\textbf{Temporal evolution of urban exponents by continent.}
    Same data as in Fig.~\ref{fig:world_hyperscaling}, now separated by continent. Colors indicate the years 1975--2020. Across regions, cities tend to move toward the upper-right part of the $(\beta,\gamma)$ plane as they develop over time.}
    \label{fig:division_by_continent}
\end{figure}

Despite these limitations, the dynamical model suggests an intriguing interpretation. The maximum of the $\gamma$--$\beta$ curve is obtained near the critical point $b=1$, where the model develops its strongest large-scale fluctuations. This raises the possibility that the systematic movement of real cities through the $(\beta,\gamma)$ plane over time (as in Fig.~\ref{fig:division_by_continent}) may be viewed, in the language of non-equilibrium statistical physics, as a drift toward increasingly correlated and possibly near-critical organization. At present we do not regard this as an established claim: the model is too stylized, and the empirical evidence is not yet sufficient to substantiate a direct connection to criticality. Nevertheless, the analogy is suggestive and points to an interesting direction for future work, namely whether the temporal evolution of urban systems can be understood as approaching a fluctuation-dominated regime analogous to critical behavior in spatial dynamical systems.

Taken together, these two models support two qualitative lessons. First, spatial correlations are indeed sufficient to generate nontrivial hyperscaling relations between $\beta$ and $\gamma$, consistent with the theoretical interpretation developed above. Second, reproducing the detailed empirical form of this relation is substantially harder than showing that a relation exists. This is because simplified models can capture the presence of correlated fluctuations while still missing important ingredients of real urban population fields, such as heterogeneous local capacities, multilevel occupancy, and realistic long-range spatial structure. For this reason, we do not present the models as explanations of the measured exponents, but rather as proof-of-principle examples showing that the hyperscaling relation uncovered in the data provides a useful and nontrivial benchmark for future urban growth models.

\section{Illustrative transport comparison and longitudinal scaling tracks: Shanghai and São Paulo}
\label{app:metro}

Transportation infrastructure is one possible source of temporal heterogeneity in urban population organization. Previous work has shown that rail and subway investments can reshape the spatial distribution of population and employment, alter urban centralization, and induce changes in land use even when their effects on aggregate population growth are limited \cite{gonzaleznavarro2018subways,jin2018seoul,heblichetal2020metropolis}. These mechanisms map naturally onto the quantities studied here. New transport links may affect the geometry of the populated support, and therefore $\beta$, through corridor development, infill, densification, or outward expansion, while changes in accessibility may reorganize the spatial distribution of population and hence the covariance structure contributing to $\gamma$. In São Paulo in particular, historical studies point to distinct and mutually dependent relationships between transport infrastructure, outward urban expansion, and densification \cite{costa2021chasing,costa2025roads}. Here we explore, as an illustrative example, whether major transportation expansions are associated with the temporal evolution of the hyperscaling relation in Shanghai and São Paulo.

\begin{figure*}[!t]
  \centering
  \includegraphics[width=0.9\textwidth]{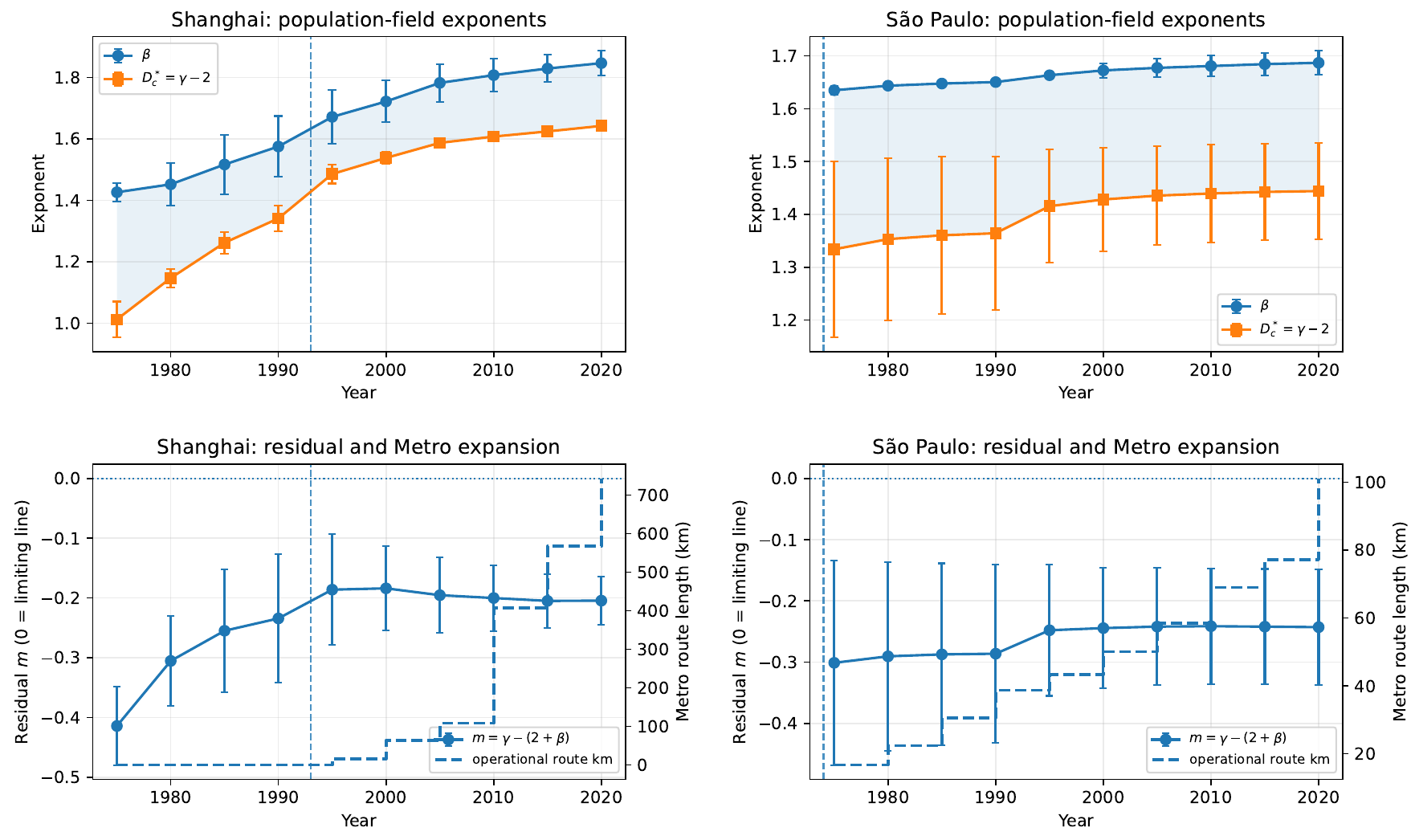}
  \caption{\textbf{Metro expansion and longitudinal hyperscaling tracks in Shanghai and São Paulo.} Top: temporal evolution of the occupied-support exponent $\beta$ and $\gamma-2$, the latter corresponding to the theory-implied $D_c^{\ast}$ only in the correlation-dominated regime. The separation between the two curves equals the residual $m=\gamma-(2+\beta)$. Bottom: $m$ together with operational metro route length. The vertical dashed lines indicate the beginning of metro operation (Shanghai, 1993; São Paulo, 1974, immediately before the first population observation). In both cities, $m$ is approximately stable from the mid-1990s onward despite continued changes in both exponents and large increases in metro route length. Error bars are approximate 95\% intervals and ignore covariance between the fitted $\beta$ and $\gamma$ estimates. The comparisons are descriptive and do not identify a causal effect of transportation infrastructure.}
  \label{fig:transport_two_city}
\end{figure*}

The two cities provide complementary histories. The Shanghai metro began operation in 1993 and subsequently underwent exceptionally rapid network expansion \cite{yang2018shanghai,shanghai2021basicfacts}, so that the population data contain four pre-opening observations (1975--1990) followed by a long period of metro growth. The São Paulo metro began operation in 1974 \cite{metrosp_architecture}, immediately before the first population observation in 1975, providing a long post-opening trajectory. Historical analyses of São Paulo also document the close co-evolution of transportation infrastructure and urban development over this period \cite{costa2021chasing,costa2025roads}. The two cases are intended as illustrative comparisons rather than as statistically representative or causal tests (see Fig.~\ref{fig:transport_two_city}).

For each city, we align the five-year population exponent series with the historical operational route length of the metro system. To measure the city's displacement from the theoretical limiting relation, we define
\begin{equation}
m=\gamma-(2+\beta).
\end{equation}
Equivalently, defining $D_c^{\ast}=\gamma-2$, we have
\begin{equation}
m=D_c^{\ast}-\beta.
\end{equation}
Here $D_c^{\ast}$ is only the value of the correlation dimension that would be implied if the correlation-dominated relation $\gamma=2+D_c$ applied over the fitted scales; it should not be interpreted as a direct empirical estimate of $D_c$. The condition $m=0$ therefore denotes the predicted limiting relation $\gamma=2+\beta$.

Shanghai provides a useful temporal contrast. Before metro service began, the residual already moved substantially toward the limiting line, from $m=-0.414$ in 1975 to $m=-0.234$ in 1990. Thus a substantial part of the long-term change in the exponent relationship preceded the start of metro operation. This observation does not exclude anticipatory effects associated with transport planning or construction, which have been documented in other settings \cite{grimaldi2024infrastructure}, but it does rule out a simple interpretation in which the observed convergence begins with metro service. After the mid-1990s, both $\beta$ and $\gamma-2$ continue to increase, but they do so at similar rates. Between 2000 and 2020, for example, $\beta$ increases from 1.722 to 1.847 and $\gamma-2$ from 1.538 to 1.642, while $m$ changes only from $-0.184$ to $-0.204$. Over the same period, operational metro route length increases from approximately 64 to 743 km, consistent with the exceptionally rapid development of the Shanghai network documented in previous studies \cite{yang2018shanghai,shanghai2021basicfacts}. Thus the rapid expansion of the metro is not accompanied by a corresponding movement toward $m=0$.

São Paulo shows a similar longitudinal stability after the mid-1990s. Across the full 1975--2020 interval, $\beta$ increases from 1.635 to 1.687 and $\gamma-2$ from 1.334 to 1.444, while operational metro route length increases from approximately 17 to 101 km. Most of the movement in $m$ occurs before the end of the 1990s. Between 2000 and 2020, approximately 51 km of route are added while $m$ changes by only $+0.0016$. As in Shanghai, the later evolution is therefore characterized primarily by coordinated changes in the two exponents rather than by a change in their relative separation. This result is compatible with the broader literature on São Paulo, which emphasizes that transportation and urban development have co-evolved through different historical phases rather than following a simple one-directional relationship \cite{costa2021chasing,costa2025roads}.

These cases highlight two related observations. First, aggregate metro expansion is not sufficient to explain convergence toward the limiting relation $\gamma=2+\beta$: large increases in route length can occur while $m$ remains approximately unchanged, and in Shanghai substantial convergence occurred before metro operation. This is consistent with broader evidence that subway expansion can alter the spatial configuration of cities without producing a proportional change in aggregate urban growth \cite{gonzaleznavarro2018subways}. Second, the longitudinal relationship between the two population exponents can itself be remarkably persistent. A nearly constant $m$ implies
\begin{equation}
\gamma \simeq \beta + \mathrm{constant},
\end{equation}
so that the city follows an approximately slope-one trajectory in the $(\beta,\gamma)$ plane. In São Paulo and Shanghai after roughly 1995, urban population structure continues to evolve, but the mean- and fluctuation-scaling exponents evolve together and preserve an approximately city-specific offset from the theoretical limiting line. The examples therefore suggest that major infrastructure expansion need not disrupt the longitudinal coupling between $\beta$ and $\gamma$, even while both exponents themselves continue to change.

The comparison does not imply that transportation has no effect on the population field. Transport infrastructure and urban development are jointly determined: infrastructure may respond to previous patterns of population and land use, while subsequent accessibility changes can in turn influence where households and firms locate \cite{costa2021chasing}. Moreover, aggregate route length does not capture where accessibility changes occur or which populations are exposed to them. Studies using spatially disaggregated data have found localized changes in population, employment, constructed area, and land use around new rail infrastructure \cite{jin2018seoul,grimaldi2024infrastructure,costa2025roads}, effects that may be averaged out in citywide exponents. Land-use responses may also be anticipatory or unfold with substantial delays \cite{grimaldi2024infrastructure}. Finally, the present analysis does not directly estimate changes in the covariance function $C(r)$ or the correlation dimension $D_c$.

A stronger test of transport-related mechanisms would therefore combine historical travel-time or ridership-weighted accessibility, station-catchment population and employment, and project-level information with direct measurements of changes in $C(r)$ or $D_2$. Such spatially resolved measures would more closely match the mechanism relevant to $\gamma$ than aggregate network size alone. The present comparison instead establishes the narrower result that metro expansion, when summarized by operational route length, does not by itself account for the observed evolution of the hyperscaling relation. At the same time, the persistence of an approximately constant offset between $\beta$ and $\gamma$ in both cities after the mid-1990s illustrates that substantial urban and infrastructural change can occur while the city-specific longitudinal coupling between the two exponents remains intact.

\onecolumn

\section{Supplementary Tables}
\label{app:tables}

\begin{longtable}{l*{7}{S[round-mode=places, round-precision=3, table-number-alignment=center]}}
\caption{\label{tab:region_result} Exponents, standard deviations and $R^2$ values supporting Figure \ref{fig:renormalizeCBS} for the Dutch urban regions in the year 2023} \\
\hline
\textbf{Region} & $ \beta$ & $\gamma$ & $\sigma_{\beta}$ &  $\sigma_{\gamma}$ & $R^{2}_{\beta}$ & $R^{2}_{\gamma}$ \\
\hline
\endfirsthead
\multicolumn{7}{c}{{\bfseries Table \thetable\ continued from previous page}} \\
\textbf{Region} & $ \beta$ & $\gamma$ & $\sigma_{\beta}$ &  $\sigma_{\gamma}$ & $R^{2}_{\beta}$ & $R^{2}_{\gamma}$ \\
\hline 
\endhead
\hline
\multicolumn{7}{|r|}{{Continued on next page}} \\
\endfoot
\endlastfoot

\begin{filecontents*}{parameters_2023.csv}
,region,bmean,bvar,stdmean,stdvar,r_mean,r_var,rmean,rvar
0,'s-Gravenhage,1.634498080991979,3.5356485476003785,0.0229570316588007,0.0309794890134942,0.9998027876314473,0.9999232355198364,0.9996056141556128,0.9998464769324582
1,'s-Hertogenbosch,1.3208693388102937,3.281196063727017,0.01131244155632,0.0370863879910179,0.9999266592748588,0.9998722733567952,0.9998533239285795,0.9997445630276858
2,Aa en Hunze,1.120286607763308,3.04302324106559,0.0260006617017239,0.0621890688847822,0.999461780059838,0.9995826065816494,0.99892384980038,0.9991653873805646
3,Aalten,0.9183562426740356,2.9083383781073606,0.0551859267203303,0.0645842027836096,0.9964083858582689,0.9995072329768054,0.9928296714086808,0.9990147087729501
4,Achtkarspelen,1.0787458460130666,2.968596660044045,0.0120051088434471,0.0815558992418255,0.9998761735963796,0.9992460942935484,0.9997523625257374,0.998492756960911
5,Alkmaar,1.317249286351579,3.349541799533001,0.0216171740902775,0.0446991735517068,0.9997307929084852,0.9998219622118966,0.9994616582894286,0.9996439561212471
6,Almelo,1.2030516651214225,3.2344067894665174,0.0175378641140543,0.062476973149099,0.9997875550403076,0.9996270868502366,0.9995751552134761,0.9992543127646905
7,Alphen aan den Rijn,1.3420602309727918,3.345037802819637,0.0107578762970606,0.0416604018035314,0.9999357508743708,0.9998449242763991,0.9998715058766917,0.9996898726012784
8,Amersfoort,1.4013145236298366,3.4545904043534743,0.0163357834676155,0.0403880405360189,0.9998641310646788,0.9998633456651016,0.9997282805897251,0.9997267100046104
9,Amsterdam,1.4585697098282822,3.349273990382959,0.0097622456330753,0.0252914091014994,0.9999552064069466,0.9999429825809436,0.9999104148203591,0.9998859684128732
10,Apeldoorn,1.1544129877643394,3.191665817566355,0.0289927887675861,0.036933893613014,0.9993698456845715,0.999866116166611,0.9987400884636043,0.9997322502581029
11,Arnhem,1.4189006372963484,3.3691144065239413,0.0085902381601985,0.0349152134474103,0.9999633492753917,0.99989261888028,0.9999266998940589,0.9997852492912649
12,Assen,1.206325681124855,3.20297157878566,0.0410061597425326,0.0708325523226057,0.9988465013032662,0.9995113013237334,0.9976943331657758,0.9990228414738631
13,Baarle-Nassau,0.8224667263056694,2.654347878648067,0.0290498036976427,0.0923775490140547,0.9987548032833288,0.9987909923890924,0.9975111570815207,0.997583446477588
14,Barneveld,1.090140481351004,3.012968316023133,0.0710925192311557,0.0235032545903366,0.9957740620601176,0.9999391546712632,0.991565982671707,0.9998783130446803
15,Bergen op Zoom,1.1734652087834243,3.139350575899684,0.0265852021151186,0.0675644540239755,0.9994871323650116,0.9995371339813138,0.9989745277632343,0.9990744822075789
16,Bodegraven-Reeuwijk,1.2858614483418689,3.2831643089415694,0.0092106652302067,0.0468260778248085,0.9999486949375492,0.9997966436029296,0.9998973925073079,0.9995933285596835
17,Borger-Odoorn,1.0969885348879138,2.566860906365371,0.0189987134217816,0.0728736021444389,0.99970018810296,0.9991949717372384,0.9994004660930936,0.9983905915449806
18,Boxtel,1.1560551516683042,3.109153608879718,0.0285374196732658,0.0537621099908441,0.9993911987185404,0.9997011360544724,0.9987827680760811,0.9994023614286027
19,Breda,1.286526834946552,3.3365448735833856,0.0156093005371273,0.0459492169855054,0.9998528252515256,0.9998103998123984,0.9997056721634578,0.9996208355730279
20,Coevorden,1.0062537153576423,2.885379924312571,0.0140196495551397,0.0740494963812361,0.9998059414018932,0.999342024974084,0.9996119204625258,0.9986844828793027
21,Culemborg,1.033573945024025,2.9934747035742286,0.0197354539680237,0.0868546637215997,0.9996356039147736,0.9991592108331244,0.9992713406140541,0.998319128592672
22,Dalfsen,0.8197787212723743,2.653568263909431,0.0334955989958523,0.0698881075507371,0.9983346853444156,0.999307061899746,0.9966721439617333,0.9986146039627029
23,De Marne,0.8909396836542389,2.6987428701113463,0.0539023338904946,0.1511972148017158,0.9963596594428702,0.996875894607331,0.9927325709651122,0.9937615492491664
24,De Wolden,0.987168860275602,2.92569512940354,0.0131095242982519,0.0755177409009926,0.9998236903416802,0.999334411084016,0.999647411768456,0.9986692651766371
25,Delfzijl,1.019721174197042,2.8881220639681446,0.0318296278380104,0.0723130387420632,0.9990271047070752,0.9993736833212916,0.9980551559394014,0.9987477589151653
26,Deventer,1.0067289584323258,2.9518415759421623,0.0207363687111585,0.0744454935862823,0.9995760018009454,0.9993645569573716,0.9991521833763636,0.9987295177026035
27,Doetinchem,1.0825274956692823,3.00161819195986,0.0419060429560381,0.0546622191915925,0.9985047956679842,0.9996685273929145,0.9970118269719629,0.9993371646599182
28,Dongeradeel,0.9357553700403148,2.818694111009839,0.0393937939912381,0.1330973103819875,0.9982324231675368,0.9977777489715932,0.9964679706629322,0.9955604363428197
29,Edam-Volendam,1.261577036855318,3.350035807525497,0.0325053402672847,0.0613431188688887,0.9993367921257555,0.9996648691199724,0.9986740240961955,0.9993298505526517
30,Ede,1.284840809680486,3.225657588091231,0.04259784970816,0.010396393123176,0.9989026090738712,0.9999896122355888,0.9978064224145872,0.9999792245790833
31,Eemsmond,1.0068418315926675,2.8825730018552864,0.0429231296120128,0.1017634927544779,0.9981874989426388,0.99875602282327,0.9963782830453606,0.9975135931257563
32,Eindhoven,1.5088959617586737,3.4756472756196577,0.0096590001807028,0.0145552869480274,0.9999590249905472,0.9999824628616896,0.9999180516600459,0.9999649260309303
33,Elburg,1.1792357714298187,3.041467820063934,0.0419739857884584,0.0552450896300067,0.9987354536677896,0.9996702338691912,0.9974725064130056,0.9993405764840834
34,Emmen,1.1461982801924484,2.752931718137752,0.0125682946184382,0.052893555919227,0.9998797860458696,0.9996310442820294,0.9997595865431339,0.9992622246923807
35,Enschede,1.3345272916865814,3.391424017306721,0.0144626741879984,0.0373422706252022,0.9998825735898956,0.999878784462732,0.999765160968753,0.9997575836186705
36,Epe,1.049670519226074,2.9339963582152717,0.0587765165896586,0.0441899116811124,0.99687920699461,0.9997732332471024,0.9937681533382025,0.999546517917365
37,Goes,1.1502211810354297,3.131680376741456,0.0377690785149935,0.1051201787956952,0.9989235140940268,0.998875177274354,0.9978481870099594,0.9977516197748723
38,Gorinchem,1.2097893757359846,3.0660135882726527,0.0111531145228628,0.0674764545629521,0.9999150198626922,0.9995160048872508,0.999830046947008,0.9990322440257707
39,Gouda,1.3524756579003074,3.3449311938590083,0.0089394648366667,0.0469937127533851,0.9999563146653462,0.9998026775133768,0.9999126312391008,0.9996053939629174
40,Groningen,1.316394186851556,3.2858113096744312,0.0286118596043888,0.0213230837482697,0.9995279232151124,0.9999578897953488,0.9990560692867156,0.999915781363967
41,Grootegast,1.0511794206867553,3.0291026685759523,0.016073377085889,0.0911447779004791,0.9997662733216828,0.9990958379490652,0.9995326012715257,0.9981924934071448
42,Haarlem,1.5367149782271474,3.4968097001309317,0.0077721824729246,0.0233903101389999,0.9999744210528576,0.9999552597076422,0.9999488427599977,0.9999105214169782
43,Hardenberg,0.9046702744539806,2.7481772242183737,0.0280765062697807,0.0679755157776815,0.9990382136940442,0.9993887523298212,0.9980773524209867,0.9987778782833566
44,Harderwijk,1.396338981826872,3.3126093230894367,0.0227291066004462,0.0209104796779128,0.9997351436187704,0.9999601561226326,0.9994703573864434,0.9999203138327997
45,Heerlen,1.4276633253332438,3.3867424919354465,0.0126699651546874,0.0337065143099573,0.9999212504770894,0.999900962695372,0.999842507155666,0.9998019351991316
46,Hellendoorn,1.003310168283345,2.978468151106956,0.0293770289812063,0.0703190932573645,0.9991437762752968,0.9994430737605422,0.9982882856696604,0.9988864576879206
47,Hellevoetsluis,1.1441354460626,3.095967057007694,0.0142248950927526,0.0668224327074389,0.99984545940498,0.9995344690557348,0.9996909426927556,0.9990691548305297
48,Helmond,1.3621960487656364,3.3943139222168357,0.0156306869675908,0.0587546328069019,0.9998683590111762,0.9997005080963188,0.9997367353517024,0.999401105888038
49,Heusden,1.2752066941304117,3.249869275915945,0.0170487640664232,0.0714884655748387,0.9998213067229956,0.9995164681511352,0.9996426453772785,0.9990331701053192
50,Hilversum,1.4395512070067735,3.4839772359093284,0.0137437268138578,0.0403760812622426,0.99990886285525,0.9998657203402894,0.9998177340164792,0.9997314587116058
51,Hoogeveen,0.9405216249562456,2.8739153446291894,0.0183459977412823,0.0576480126328569,0.9996197253995928,0.9995978771828457,0.9992395954079573,0.9991959160684514
52,Hoorn,1.3273699027020207,3.329043561111636,0.0131563892003676,0.055541604565991,0.9999017742166534,0.9997217618975176,0.9998035580816113,0.9994436012114769
53,Hulst,0.9522842521369412,2.817626835774746,0.035739730468524,0.1532919285724306,0.9985944285036878,0.9970532135790572,0.997190832638607,0.994115110708325
54,Kollumerland en Nieuwkruisland,1.0159115984547522,2.915229133107803,0.0244281126005589,0.1210847159053752,0.999422314410878,0.9982792761208455,0.998844962542396,0.9965615131323594
55,Leek,1.0553501289682117,3.0369514443774617,0.0138579656996702,0.0818842496024808,0.999827617362802,0.9992738074693952,0.9996552644413776,0.9985481422943819
56,Leerdam,0.9913705170684256,2.8081664909813684,0.0108036173600099,0.0876857444141905,0.9998812621901708,0.9990264064652964,0.9997625384790091,0.9980537608149636
57,Leeuwarden,1.085030869799727,3.053131568402487,0.0439594944523694,0.0748874695960276,0.9983626048398204,0.9993989154514064,0.9967278907425514,0.9987981922054473
58,Leiden,1.550788317578624,3.497327453332833,0.0064090768205065,0.0068401217940589,0.9999829205143992,0.9999961748137712,0.9999658413205073,0.9999923496421745
59,Leudal,1.1665110806101806,3.149418660126046,0.0143791074425305,0.0692724791757953,0.9998480897245472,0.9995165565897112,0.9996962025258261,0.9990333468969533
60,Loppersum,0.9732571940304976,2.907707498630742,0.035312318019104,0.1011677133478091,0.9986861654881416,0.9987916445465173,0.9973740571374078,0.9975847492159364
61,Maasgouw,1.1834607722024693,3.148480089927896,0.0267812008170005,0.0800974925408754,0.9994882958809076,0.9993534314470456,0.9989768536029207,0.9987072809449848
62,Maastricht,1.4111717536639796,3.2656671443453966,0.0072786287478434,0.0186923705473789,0.999973397524204,0.999967238514402,0.9999467957560997,0.9999344781021189
63,Meppel,1.0803069326197212,3.0244597595835865,0.0146051059980988,0.0866543860958195,0.9998172758065814,0.9991801184417696,0.9996345850012937,0.9983609090893086
64,Middelburg,1.251927191052786,3.288421445127681,0.0366037794272198,0.0727224938369848,0.9991462371346512,0.9995112979825724,0.9982932031803327,0.9990228347948067
65,Nijmegen,1.418573423343097,3.3497037569654293,0.0077987960631779,0.0116671324108504,0.9999697774015756,0.9999878686942136,0.9999395557165566,0.9999757375355959
66,Noordoostpolder,0.8080278583578987,2.7047686763856507,0.0441262511199654,0.104329737948929,0.9970310428018474,0.9985154731549484,0.9940709003105392,0.9970331501298504
67,Noordwijkerhout,1.3573087353159872,3.3473503599796084,0.0279436781714439,0.0590179976083444,0.9995764215579732,0.9996892836205176,0.999153022534643,0.9993786637857036
68,Ommen,0.9235052829513436,2.7544219660262046,0.0357477391398537,0.0628522902158509,0.9985049916070646,0.9994797138932122,0.9970122182642243,0.9989596984840573
69,Oss,1.221947127365029,3.207771690109476,0.0051438420466918,0.0584113122584072,0.9999822801975752,0.9996685858195002,0.9999645607091417,0.9993372814743595
70,Purmerend,1.267685244873204,3.3636665645935238,0.0370685459581111,0.0537758360019735,0.9991460515474204,0.9997445052385132,0.9982928323228005,0.9994890757545996
71,Raalte,0.8598398223837472,2.763292148173853,0.0291442086155533,0.0660649745149974,0.9988531099327428,0.9994288939362984,0.9977075352223119,0.9988581140347328
72,Rijssen-Holten,0.9084899937168,2.9267661022117,0.0279581086279997,0.0841803038002066,0.9990542880799144,0.9991737595787824,0.9981094705308645,0.9983482018307984
73,Roermond,1.2836679051685542,3.252802302016928,0.0211628007752701,0.0626994648909696,0.999728316097428,0.9996286611625808,0.9994567060069989,0.9992574602176937
74,Roosendaal,1.215134809507328,3.214616289145772,0.0205289708878765,0.0554809125987579,0.9997147014613348,0.9997022615540616,0.9994294843179258,0.9994046117563055
75,Rotterdam,1.471296575706709,3.3437085377935505,0.0070603586279561,0.0224713451806599,0.9999769729622218,0.9999548381324666,0.999953946454688,0.9999096783045274
76,Schouwen-Duiveland,0.9991080231619563,2.889192558407589,0.0521118496464318,0.1412491114049879,0.9972905556013064,0.997618421751972,0.9945884522915625,0.9952425154188955
77,Sittard-Geleen,1.4152925101949287,3.4183782683659443,0.0189480609725269,0.0506103011743245,0.9998208072705846,0.999780873531622,0.9996416466512036,0.9995617950796533
78,Sluis,0.818839611692765,2.6657866525990994,0.0488472444595688,0.1548047639589785,0.9964602569148744,0.9966447273882612,0.9929330436108574,0.9933007126308214
79,Smallingerland,1.077586234153859,3.041984048457085,0.0173462666550809,0.075712813921548,0.9997409764788742,0.9993810984276588,0.9994820200509328,0.9987625798944738
80,Stadskanaal,1.043588677623933,2.8061706022222923,0.0189270509130645,0.0342683007100834,0.9996712294050388,0.9998509061042196,0.9993425669001818,0.9997018344374289
81,Staphorst,1.0078310808776911,2.9310561589031026,0.0107709148476584,0.083501802225336,0.999885802843011,0.9991893841337276,0.9997716187270126,0.9983794253655378
82,Stede Broec,1.2678950777001206,3.3243305793653457,0.0298934688335868,0.1034323171553983,0.9994445767541482,0.9990333393928914,0.9988894620032784,0.9980676132185121
83,Terneuzen,0.9981377809688736,2.908968504808693,0.0538346998831784,0.1114551104880707,0.9971036329734928,0.9985352385142786,0.9942156548889378,0.9970726225547673
84,Tholen,1.0327302374569245,2.977048741098299,0.0448625581054569,0.1435191767081389,0.9981182277089676,0.9976840072124104,0.9962399964848906,0.9953733782474129
85,Tiel,1.106817555854281,3.017973674090594,0.0125626243932047,0.0735652671566037,0.9998711973618843,0.999406353544996,0.9997424113138883,0.9988130595061054
86,Tilburg,1.3677143955417967,3.433977024277272,0.0142131903748298,0.0364405144399331,0.9998920253098176,0.9998874097255526,0.9997840622781689,0.999774832127675
87,Uitgeest,1.4618060393695138,3.4647569964858995,0.0153527487820272,0.0345531410811041,0.9998897138607116,0.999900559174186,0.9997794398844556,0.9998011282368497
88,Uithoorn,1.2925830829174985,3.1804655734812903,0.0128057445851523,0.0435498206570617,0.9999018638983928,0.999812557001819,0.99980373742748,0.9996251491385156
89,Utrecht,1.439003694512207,3.395705972563458,0.0117369464180615,0.0130456237453753,0.9999334814013586,0.9999852408982912,0.9998669672274412,0.9999704820144135
90,Veendam,1.1265951213872671,3.0275580869494134,0.0221435255632656,0.0578876818221626,0.9996138942927062,0.9996346162147602,0.9992279376630296,0.999269365934831
91,Veenendaal,1.3217714352418053,3.3499255572111557,0.038737726303759,0.0183656268679121,0.9991421794842762,0.9999699446616678,0.9982850948245896,0.999939890226659
92,Venlo,1.2021274136287858,3.173072910259934,0.0134867745545856,0.0489875684431937,0.9998741557990553,0.999761737097592,0.9997483274348734,0.9995235309643946
93,Waadhoeke,0.9923529832603676,2.9128139358173364,0.045705381171678,0.1035544897784587,0.9978854250307928,0.998738492099402,0.995775321488886,0.9974785756009873
94,Waalwijk,1.280001220566104,3.331650631286757,0.0247751680177143,0.0867742239761357,0.9996255718622724,0.999322325833701,0.9992512839209752,0.9986451109096776
95,Weert,1.198884411459514,3.1632631500201733,0.0084012605449122,0.040853978036265,0.9999508976483518,0.9998332409495456,0.9999017977077445,0.9996665097076721
96,West Maas en Waal,1.1615931580987526,3.060065381070449,0.006740449165508,0.0725486828071249,0.999966329648509,0.9994383940778192,0.9999326604307105,0.9988771035568502
97,Westerveld,0.9406805245130994,2.8358746899403973,0.0136592327150078,0.0740777141042355,0.999789219235711,0.9993183572114748,0.9995784828999525,0.9986371790598407
98,Weststellingwerf,0.9641722893414368,2.9044217828595484,0.0161796364517161,0.0742678830438667,0.9997185217578692,0.9993467836831724,0.9994371227457393,0.9986939940579014
99,Wijchen,1.2837799084905224,3.2056975634428806,0.0097182660114291,0.0169028053323953,0.9999426994109312,0.9999721993868468,0.9998854021052198,0.9999443995465677
100,Wijk bij Duurstede,1.0754368314923488,3.087292183231184,0.0260652120788735,0.1074482857219513,0.9994130917069952,0.9987909192309122,0.9988265278753349,0.9975833003381307
101,Winterswijk,0.8574559137571571,2.865215314351935,0.0261333262354143,0.0787837712417912,0.9990723990709248,0.9992447915977865,0.9981456585853332,0.9984901535353037
102,Woerden,1.1667070009750793,3.157359536232738,0.0168368333922922,0.0689902641131065,0.9997918091778252,0.9995228920614724,0.9995836616990688,0.9990460117549298
103,Zaanstad,1.4029346511282763,3.2816196744827533,0.0245257291974081,0.021314390211287,0.9996945290318836,0.99995781657385,0.9993891513762796,0.9999156349271414
104,Zeist,1.4028345915540652,3.2389220488102493,0.0096686368978644,0.0164814438884319,0.9999525007639152,0.9999741075794164,0.9999050037840078,0.9999482158292502
105,Zevenaar,1.2340052787014348,3.2308737814770496,0.0085446309125389,0.0677987506553829,0.9999520574258768,0.9995599351833908,0.999904117150244,0.9991200640238245
106,Zuidhorn,0.9999829929289784,2.8609171912078093,0.0434006496079399,0.0575225212997961,0.9981216252659416,0.9995959808083784,0.9962467788235247,0.999192124848264
107,Zutphen,1.0004308838504192,2.949263339072879,0.0218034141357362,0.0673286861958565,0.9995253586827216,0.9994792445419836,0.9990509426498233,0.9989587602702141
108,Zwolle,1.129491558466086,3.1040512277335006,0.0192303079149402,0.0643842015853516,0.999710253824093,0.9995700472713764,0.9994205916010324,0.9991402794021017
\end{filecontents*}
\csvreader[head to column names]{parameters_2023.csv}{} 
{ \region & \bmean & \bvar & \stdmean &  \stdvar & \rmean & \rvar \\[1pt] }

\end{longtable}

\def\prevcontinent{}

{\small
\begin{longtable}{@{}p{0.22\linewidth}p{0.74\linewidth}@{}}
\caption{List of cities per continent used for the analysis of the GHS-POP dataset.}
\label{tab:continent_regions}\\
\toprule
\textbf{Continent} & \textbf{City} \\
\midrule
\endfirsthead

\toprule
\textbf{Continent} & \textbf{City} \\
\midrule
\endhead

\midrule
\multicolumn{2}{r}{\scriptsize Continued on next page}\\
\endfoot

\bottomrule
\endlastfoot


\textbf{Africa} & Abidjan (Ivory Coast), Abuja (Nigeria), Accra (Ghana), Addis Ababa (Ethiopia), Alexandria (Egypt), Algiers (Algeria), Antananarivo (Madagascar), Arusha (Tanzania), Asmara (Eritrea), Bamako (Mali), Bangui (Central African Republic), Bata (Equatorial Guinea), Beira (Mozambique), Blantyre (Malawi), Bobo-Dioulasso (Burkina Faso), Bouaké (Ivory Coast), Bujumbura (Burundi), Bulawayo (Zimbabwe), Butare (Rwanda), Cairo (Egypt), Cape Town (South Africa), Casablanca (Morocco), Conakry (Guinea), Constantine (Algeria), Cotonou (Benin), Dakar (Senegal), Dar es Salaam (Tanzania), Dire Dawa (Ethiopia), Djibouti (Djibouti), Dodoma (Tanzania), Douala (Cameroon), Durban (South Africa), Fes (Morocco), Francistown (Botswana), Gaborone (Botswana), Gitega (Burundi), Goma (Democratic Republic of the Congo), Gulu (Uganda), Harare (Zimbabwe), Ibadan (Nigeria), Johannesburg (South Africa), Kampala (Uganda), Kano (Nigeria), Khartoum (Sudan), Kigali (Rwanda), Kinshasa (Democratic Republic of the Congo), Kisumu (Kenya), Kumasi (Ghana), Lagos (Nigeria), Libreville (Gabon), Lilongwe (Malawi), Lobito (Angola), Luanda (Angola), Lubumbashi (Democratic Republic of the Congo), Lusaka (Zambia), Malabo (Equatorial Guinea), Maputo (Mozambique), Marrakech (Morocco), Mbabane (Eswatini), Mombasa (Kenya), Moroni (Comoros), Moundou (Chad), N'Djamena (Chad), Nairobi (Kenya), Nampula (Mozambique), Ndola (Zambia), Oran (Algeria), Ouagadougou (Burkina Faso), Pointe-Noire (Republic of the Congo), Port Harcourt (Nigeria), Port-Gentil (Gabon), Porto-Novo (Benin), Praia (Cape Verde), Pretoria (South Africa), Rabat (Morocco), Serrekunda (Gambia), Sfax (Tunisia), Sousse (Tunisia), Tamale (Ghana), Toamasina (Madagascar), Touba (Senegal), Tunis (Tunisia), Walvis Bay (Namibia), Windhoek (Namibia), Yamoussoukro (Ivory Coast), Yaoundé (Cameroon) \\ \hline
\textbf{Asia} & Abu Dhabi (United Arab Emirates), Ahmedabad (India), Aleppo (Syria), Ankara (Turkey), Baghdad (Iraq), Bandung (Indonesia), Bangalore (India), Bangkok (Thailand), Basra (Iraq), Beijing (China), Bursa (Turkey), Busan (South Korea), Cebu (Philippines), Chengdu (China), Chennai (India), Chiang Mai (Thailand), Chittagong (Bangladesh), Chongqing (China), Da Nang (Vietnam), Damascus (Syria), Delhi (India), Dhaka (Bangladesh), Dubai (United Arab Emirates), Erbil (Iraq), Faisalabad (Pakistan), George Town (Malaysia), Guangzhou (China), Haifa (Israel), Hanoi (Vietnam), Ho Chi Minh City (Vietnam), Hyderabad (India), Isfahan (Iran), Islamabad (Pakistan), Istanbul (Turkey), Izmir (Turkey), Jaipur (India), Jakarta (Indonesia), Jeddah (Saudi Arabia), Jerusalem (Israel), Johor Bahru (Malaysia), Kabul (Afghanistan), Karachi (Pakistan), Khulna (Bangladesh), Kolkata (India), Kuala Lumpur (Malaysia), Kurume (Japan), Lahore (Pakistan), Manila (Philippines), Mashhad (Iran), Mecca (Saudi Arabia), Medan (Indonesia), Medina (Saudi Arabia), Mumbai (India), Nagoya (Japan), Nanjing (China), Osaka (Japan), Pune (India), Quezon City (Philippines), Riyadh (Saudi Arabia), Samarkand (Uzbekistan), Sapporo (Japan), Semarang (Indonesia), Seoul (South Korea), Shanghai (China), Shenzhen (China), Singapore (Singapore), Surabaya (Indonesia), Surat (India), Tabriz (Iran), Taipei (Taiwan), Tashkent (Uzbekistan), Tehran (Iran), Tel Aviv (Israel), Tianjin (China), Tokyo (Japan), Wuhan (China), Xi'an (China) \\ \hline
\textbf{Europe} & Aarhus (Denmark), Amsterdam (the Netherlands), Antwerp (Belgium), Athens (Greece), Barcelona (Spain), Basel (Switzerland), Belgrade (Serbia), Bergen (Norway), Berlin (Germany), Bilbao (Spain), Birmingham (United Kingdom), Braga (Portugal), Bratislava (Slovakia), Brno (Czech Republic), Brussels (Belgium), Bucharest (Romania), Budapest (Hungary), Cluj-Napoca (Romania), Cologne (Germany), Copenhagen (Denmark), Cork (Ireland), Debrecen (Hungary), Dublin (Ireland), Edinburgh (United Kingdom), Frankfurt (Germany), Gdansk (Poland), Geneva (Switzerland), Ghent (Belgium), Glasgow (United Kingdom), Gothenburg (Sweden), Graz (Austria), Hamburg (Germany), Helsinki (Finland), Kharkiv (Ukraine), Košice (Slovakia), Krakow (Poland), Kyiv (Ukraine), Limerick (Ireland), Linz (Austria), Lisbon (Portugal), Ljubljana (Slovenia), London (United Kingdom), Lyon (France), Madrid (Spain), Malmö (Sweden), Manchester (United Kingdom), Maribor (Slovenia), Marseille (France), Milan (Italy), Moscow (Russia), Munich (Germany), Naples (Italy), Nice (France), Nizhny Novgorod (Russia), Niš (Serbia), Novi Sad (Serbia), Odense (Denmark), Odesa (Ukraine), Oslo (Norway), Ostrava (Czech Republic), Palermo (Italy), Paris (France), Patras (Greece), Plovdiv (Bulgaria), Porto (Portugal), Prague (Czech Republic), Riga (Latvia), Rijeka (Croatia), Rome (Italy), Rotterdam (the Netherlands), Seville (Spain), Sofia (Bulgaria), Split (Croatia), St. Petersburg (Russia), Stavanger (Norway), Stockholm (Sweden), Szeged (Hungary), Tallinn (Estonia), Tampere (Finland), Tartu (Estonia), Thessaloniki (Greece), Timisoara (Romania), Toulouse (France), Turin (Italy), Utrecht (the Netherlands), Valencia (Spain), Vienna (Austria), Warsaw (Poland), Wroclaw (Poland), Zagreb (Croatia), Zurich (Switzerland)\\ \hline
\textbf{North and Central America} & Austin (USA), Baltimore (USA), Basseterre (Saint Kitts and Nevis), Belmopan (Belize), Boston (USA), Bridgetown (Barbados), Calgary (Canada), Castries (St. Lucia), Charlotte (USA), Chicago (USA), Ciudad Juárez (Mexico), Columbus (USA), Dallas (USA), Denver (USA), Detroit (USA), Edmonton (Canada), El Paso (USA), Fort Worth (USA), Guadalajara (Mexico), Guatemala City (Guatemala), Hamilton (Canada), Havana (Cuba), Houston (USA), Indianapolis (USA), Jacksonville (USA), Kingston (Jamaica), Kitchener (Canada), Las Vegas (USA), León (Mexico), Los Angeles (USA), Louisville (USA), Managua (Nicaragua), Memphis (USA), Mexico City (Mexico), Monterrey (Mexico), Montreal (Canada), Nashville (USA), Nassau (The Bahamas), New York City (USA), Oklahoma City (USA), Ottawa (Canada), Panama City (Panama), Philadelphia (USA), Phoenix (USA), Port-au-Prince (Haiti), Portland (USA), Puebla (Mexico), Quebec City (Canada), San Antonio (USA), San Diego (USA), San Francisco (USA), San José (Costa Rica), San Juan (Puerto Rico), San Salvador (El Salvador), Santo Domingo (Dominican Republic), Seattle (USA), Tegucigalpa (Honduras), Tijuana (Mexico), Toronto (Canada), Vancouver (Canada), Washington (D.C.), Winnipeg (Canada), Zapopan (Mexico) \\ \hline
\textbf{South America} & Arequipa (Peru), Asunción (Paraguay), Barranquilla (Colombia), Belo Horizonte (Brazil), Bogotá (Colombia), Buenos Aires (Argentina), Cali (Colombia), Caracas (Venezuela), Cartagena (Colombia), Ciudad del Este (Paraguay), Cochabamba (Bolivia), Concepción (Chile), Cuenca (Ecuador), Curitiba (Brazil), Córdoba (Argentina), Fortaleza (Brazil), Guayaquil (Ecuador), La Paz (Bolivia), La Plata (Argentina), Lima (Peru), Manaus (Brazil), Maracaibo (Venezuela), Medellín (Colombia), Mendoza (Argentina), Montevideo (Uruguay), Porto Alegre (Brazil), Quito (Ecuador), Recife (Brazil), Rio de Janeiro (Brazil), Rosario (Argentina), Salvador (Brazil), Santa Cruz de la Sierra (Bolivia), Santiago (Chile), São Paulo (Brazil), Trujillo (Peru), Valencia (Venezuela) \\ \hline
\textbf{Oceania} & Adelaide (Australia), Auckland (New Zealand), Brisbane (Australia), Canberra (Australia), Christchurch (New Zealand), Hobart (Australia), Honiara (Solomon Islands), Lae (Papua New Guinea), Melbourne (Australia), Newcastle (Australia), Perth (Australia), Port Moresby (Papua New Guinea), Suva (Fiji), Sydney (Australia), Wellington (New Zealand)

\end{longtable}

}

\end{appendices}

\end{document}